\documentclass[12pt]{article}

\input{epsf}
\def\slash#1{\ooalign{$\hfil/\hfil$\crcr$#1$}} 

\def\lra{\leftrightarrow}
\def\Ph{{\hat P}}
\def\hs{\hspace{.1mm}}

\newcommand\as{\alpha_{\mathrm{S}}}

\newcommand\f[2]{\frac{#1}{#2}}
\def\ep{\epsilon}
\def\ktil{\widetilde k}

\def\beq{\begin{equation}}
\def\eeq{\end{equation}}
\def\beeq{\begin{eqnarray}}
\def\eeeq{\end{eqnarray}}
\def\cm{{\cal M}}
\def\bom#1{{\mbox{\boldmath $#1$}}}
\def\to{\rightarrow}
\def\kper{k_{\perp}}

\newcommand{\la}{\langle}
\newcommand{\ra}{\rangle}

\def\nn{\nonumber}

\def\ID{1 \kern -.45 em 1}

\def\ket#1{|{#1}\ra}
\def\bra#1{\la{#1}|}

\begin{document}

\begin{titlepage}
\renewcommand{\thefootnote}{\fnsymbol{footnote}}
\begin{flushright}
     CERN--TH/99-263\\ ETH--TH/99-22\\ hep-ph/9908523
     \end{flushright}
\par \vspace{10mm}
\begin{center}
{\Large \bf
Infrared factorization of tree-level QCD amplitudes\\[1ex]
at the  next-to-next-to-leading order and 
beyond~\footnote{This work was supported in part 
by the EU Fourth Framework Programme ``Training and Mobility of Researchers'', 
Network ``Quantum Chromodynamics and the Deep Structure of
Elementary Particles'', contract FMRX--CT98--0194 (DG 12 -- MIHT).}}
\end{center}
\par \vspace{2mm}
\begin{center}
{\bf Stefano Catani}~\footnote{On leave of absence from INFN,
Sezione di Firenze, Florence, Italy.}\\

\vspace{5mm}

{Theory Division, CERN, CH 1211 Geneva 23, Switzerland} \\

\vspace{5mm}

{and}

\vspace{3mm}

{\bf Massimiliano Grazzini}~\footnote{Work supported in part by the Swiss
  National Foundation.}\\

\vspace{5mm}
Institute for Theoretical Physics, ETH-H\"onggerberg, CH 8093 Zurich, 
Switzerland

\vspace{5mm}

\end{center}

\par \vspace{2mm}
\begin{center} {\large \bf Abstract} \end{center}
\begin{quote}
\pretolerance 10000

We study the infrared behaviour of tree-level QCD amplitudes
and we derive infrared-factorization formulae that are valid at
any perturbative order. We explicitly compute all the universal infrared 
factors that control the singularities in the various 
soft and/or collinear limits at ${\cal O}(\as^2)$.

\end{quote}

\vspace*{\fill}
\begin{flushleft}
     CERN--TH/99-263 \\ ETH--TH/99-22 \\August 1999 

\end{flushleft}
\end{titlepage}

\renewcommand{\thefootnote}{\fnsymbol{footnote}}
\section{Introduction}
\label{intro}

Infrared (soft and collinear) singularities appear in the calculation
of multiparton QCD matrix elements.
Although the singularities cancel in the evaluation of inclusive cross
sections, their factorization properties are at the basis
of many important tools in perturbative QCD applications to
hard-scattering processes~[\ref{book}].

At the {\em leading} order in the QCD coupling, $\as$,
the structure of the infrared singularities is 
well known to be
universal.
It is embodied in process-independent factorization formulae
of tree-level [\ref{AP}--\ref{antenna}] and one-loop [\ref{GG}--\ref{BDKrev}] 
amplitudes. These factorization formulae have played an essential r\^ole
in the setting up of completely general algorithms
[\ref{CSdipole},\ref{GG},\ref{GGK},\ref{submeth}]
to handle and cancel infrared singularities, when
combining tree-level and one-loop contributions in the evaluation of jet cross
sections at the next-to-leading order (NLO) in perturbation theory.

The extension of these general algorithms at the next-to-next-to-leading
order (NNLO) is at present one of the main goals to improve and precisely
quantify the theoretical accuracy of perturbative QCD predictions.
To this purpose we need to compute two-loop matrix elements 
[\ref{gonsalves}--\ref{bern}] and to understand the structure of the infrared
singularities of two-loop, one-loop and tree-level amplitudes
at ${\cal O}(\as^2)$.
The singular behaviour of two-loop QCD amplitudes has been 
discussed in Ref.~[\ref{sing2loop}].
The soft and collinear limits of one-loop amplitudes have been derived in 
Refs.~[\ref{1loopeps}, \ref{1loopepskos}]. 
The soft, collinear and  soft--collinear singularities of tree-level amplitudes
have been studied in Refs.~[\ref{bgdsoft},\ref{sdsoft}],
[\ref{glover},\ref{lett}] and [\ref{glover}], respectively.

The purpose of this paper is twofold. We consider {\em tree-level} 
matrix elements and present general techniques to compute their infrared 
singularities and to derive infrared-factorization formulae to {\em any}
order in $\as$. We apply these techniques to the explicit calculation
of all the relevant infrared factors at ${\cal O}(\as^2)$.

Our general method exploits the universality properties of soft 
and collinear emission and consists in directly computing
{\em process-independent} Feynman subgraphs in a physical gauge. We use
power-counting arguments [\ref{collpc},\ref{jetcalc}] and the eikonal
approximation [\ref{BCM}] to treat the collinear and soft limits, respectively.
We show how the coherence properties of QCD radiation [\ref{coher}]
can be used to
deal with the mixed soft--collinear limit in terms of the collinear and soft
factorization formulae.

Most of the explicit results at ${\cal O}(\as^2)$ presented in this paper
were first obtained by Campbell and Glover [\ref{glover}]. The strategy
followed in Ref.~[\ref{glover}] was to take universal factorization for granted
and thus to extract the ${\cal O}(\as^2)$-singular factors by performing the
corresponding limits of a set of known matrix elements. We confirm their
calculations by using a completely independent method. We also extend their 
results by considering the emission of a soft fermion pair and
by fully taking into account spin (azimuthal) correlations in the collinear
limit. The extension to azimuthal correlations is essential to
apply some general methods to perform exact NNLO calculations of jet cross
sections. For instance,
the subtraction method [\ref{CSdipole}, \ref{submeth}] works by regularizing
the infrared singularities of the tree-level matrix element by identifying
and subtracting a proper {\em local} counterterm. Thus, the study
of the azimuthally {\em averaged} collinear limit [\ref{glover}]
is not sufficient for this purpose.

The knowledge of the infrared structure of multiparton amplitudes is also 
important for other perturbative QCD applications.
The leading-logarithmic (LL) parton showers, which are implemented in Monte 
Carlo event generators [\ref{book}] to describe the exclusive structure of
hadronic final states, are based on the ${\cal O}(\as)$-factorization
formulae supplemented with `jet calculus' techniques [\ref{KUV}] and 
colour-coherence properties [\ref{BCM}, \ref{coher}]. 
Analytical techniques
to perform all-order resummation of logarithmically enhanced contributions
at next-to-leading logarithmic (NLL) accuracy [\ref{softrev}] rely on the
factorization properties of soft and collinear emission.
The results on infrared factorization presented in this paper can be useful
to improve parton-shower algorithms and resummed calculations beyond their
present logarithmic accuracy.

The outline of the paper is as follows. We start in Sect.~\ref{seccoll} by
studying the collinear behaviour. After reviewing the known
factorization formulae at ${\cal O}(\as)$, we discuss the kinematics of the 
triple collinear limit. Then, in Sect.~\ref{power},
we present our derivation of factorization for the multiple collinear limit 
at any perturbative order. Finally, in Sect.~\ref{splitt}, we perform the 
explicit calculation of the spin-dependent splitting functions at 
${\cal O}(\as^2)$.
Our results for the splitting functions were anticipated in  Ref.~[\ref{lett}].
In Sect.~\ref{secsoft} we study the soft behaviour. 
We first review the known results at ${\cal O}(\as)$ and then, in 
Sect.~\ref{secsoftqq}, we compute the emission of a soft $q{\bar q}$ pair
at ${\cal O}(\as^2)$. Section~\ref{secsoftgg} is devoted to double gluon
emission: we present the corresponding soft current and
obtain a compact expression for its square. Factorization for the 
mixed soft--collinear limit at ${\cal O}(\as^2)$ and at higher perturbative
orders is discussed in detail in Sects.~\ref{softcoll} and \ref{multilim}.
In Sect.~\ref{summa} we summarize our results. In general,
soft factorization formulae involve colour correlations. At ${\cal O}(\as^2)$
these correlations cancel in four- and five-parton matrix elements. The
explicit expressions for these particular cases are given in the Appendix.

\section{The collinear behaviour}
\label{seccoll}
\subsection{Notation and collinear factorization at ${\cal O}(\as)$}
\label{notations}

We consider a generic scattering process involving 
final-state\footnote{The case of incoming partons can be recovered by
simply crossing the parton indices (flavours, spins and colours) and momenta.}
QCD partons ({\em massless} quarks and gluons) with momenta 
$p_1, p_2, \dots$. Non-QCD partons $(\gamma^*, Z^0, W^\pm, \dots)$, carrying
a total momentum $Q$, are always understood. The corresponding {\em tree-level}
matrix element is denoted by
\beq
\label{meldef}
\cm^{c_1,c_2,\dots;s_1,s_2,\dots}_{a_1,a_2,\dots}(p_1,p_2,\dots) \;\;,
\eeq
where $\{c_1,c_2,\dots\}$, $\{s_1,s_2,\dots\}$ and $\{a_1,a_2,\dots\}$ are
respectively colour, spin and flavour indices. The matrix element squared,
summed over final-state colours and spins, will be denoted by 
$| \cm_{a_1,a_2,\dots}(p_1,p_2,\dots) |^2$.
If the sum over the spin polarizations of 
the parton $a_1$ is not carried out, we define the following
`spin-polarization tensor'
\beq
\label{melspindef}
{\cal T}_{a_1,\dots}^{s_1 s'_1}(p_1,\dots) \equiv 
\sum_{{\rm spins} \,\neq s_1,s'_1} \, \sum_{{\rm colours}}
\cm^{c_1,c_2,\dots;s_1,s_2,\dots}_{a_1,a_2,\dots}(p_1,p_2,\dots) \,
\left[ \cm^{c_1,c_2,\dots;s'_1,s_2,\dots}_{a_1,a_2,\dots}(p_1,p_2,\dots)
\right]^\dagger
\;\;.
\eeq

We work in $d=4 - 2\ep$ space-time
dimensions and consider two helicity states for massless quarks and 
$d-2$ helicity states for gluons. This defines the conventional 
dimensional-regularization (CDR) scheme of both ultraviolet [\ref{cdruv}]
and infrared [\ref{cdrir}] divergences.
Thus, the fermion spin indices are $s=\pm 1$,
while to label the gluon spin it is convenient to use the corresponding Lorentz
index $\mu=1, \dots, d$. The $d$-dimensional average of the matrix element
over the polarizations of a parton $a$ is obtained by means of the factors
\beq
\label{ferav}
\frac{1}{2} \delta_{ss'}
\eeq
for a fermion, and (the gauge terms are proportional either to $p^\mu$
or to $p^\nu$)
\beq
\label{gluav}
\frac{1}{d-2} d_{\mu \nu}(p) = \frac{1}{2(1-\ep)} ( - g_{\mu \nu} +
{\rm gauge \; terms} )
\eeq
with 
\beq
\label{dprop}
- g^{\mu \nu} d_{\mu \nu}(p) = d-2 \;, \;\;\;\;\;
p^\mu \,d_{\mu \nu}(p) = d_{\mu \nu}(p) \,p^\nu = 0 \;,
\eeq
for a gluon with on-shell momentum $p$.

The 
singular collinear limit at ${\cal O}(\as)$ is approached when the momenta
of two partons, say $p_1$ and $p_2$, become parallel. This limit can
be precisely defined as follows:
\beeq
\label{clim}
&&p_1^\mu = z p^\mu + k_\perp^\mu - \frac{k_\perp^2}{z} 
\frac{n^\mu}{2 p\cdot n} \;\;, \;\;\; p_2^\mu =
(1-z) p^\mu - k_\perp^\mu - \frac{k_\perp^2}{1-z} \frac{n^\mu}{2 p\cdot n}\;\;,
\nonumber \\
&&s_{12} \equiv 2 p_1 \cdot p_2 = - \frac{k_\perp^2}{z(1-z)} \;\;,
\;\;\;\;\;\;\;\; k_\perp \to 0 \;\;.
\eeeq
In Eq.~(\ref{clim}) the light-like ($p^2=0$) vector $p^\mu$ denotes the
collinear direction, while $n^\mu$ is an auxiliary light-like vector, which 
is necessary to specify the transverse component $k_\perp$ ($k_\perp^2<0$)
($k_\perp \cdot p = k_\perp \cdot n = 0$) or, equivalently, how the collinear 
direction is approached.
In the small-$k_\perp$ limit (i.e.\ neglecting terms that are less singular 
than $1/k_\perp^2$), the square of the matrix element in Eq.~(\ref{meldef})
fulfils the following factorization formula [\ref{book}]:
\beeq
\label{cfac}
| \cm_{a_1,a_2,\dots}(p_1,p_2,\dots) |^2 \simeq \frac{2}{s_{12}} \;
4 \pi \mu^{2\ep} \as 
\;{\cal T}_{a,\dots}^{s s'}(p,\dots) \;
{\hat P}_{a_1 a_2}^{s s'}(z,\kper;\ep) \;\;,
\eeeq
where $\mu$ is the dimensional-regularization scale.
The spin-polarization tensor ${\cal T}_{a,\dots}^{s s'}(p,\dots)$
is obtained by replacing the partons $a_1$ and $a_2$ on the right-hand side
of Eq.~(\ref{melspindef}) with a single parton denoted by $a$.
This parton carries the quantum numbers of the
pair $a_1+a_2$ in the collinear limit. In other words, its momentum is 
$p^\mu$ and its other quantum numbers (flavour, colour) are obtained according
to the following rule: anything~+~gluon gives anything, and 
quark~+~antiquark gives gluon.

The kernel ${\hat P}_{a_1 a_2}$ in Eq.~(\ref{cfac}) is the $d$-dimensional
Altarelli--Parisi splitting function [\ref{AP}]. 
It depends not only on the momentum
fraction $z$ involved in the collinear splitting $a \to a_1 + a_2$, but also on
the transverse momentum $\kper$ and on the helicity of the parton $a$ in the
matrix element $\cm_{a,\dots}^{c,\dots;s,\dots}(p,\dots)$.
More precisely, ${\hat P}_{a_1 a_2}$ is in general a matrix
acting on the spin indices $s,s'$ of the parton $a$ in the 
spin-polarization tensor ${\cal T}_{a,\dots}^{s s'}(p,\dots)$.
Because of these {\em spin correlations}, the spin-average square
of the matrix element $\cm_{a,\dots}^{c,\dots;s,\dots}(p,\dots)$
cannot be simply factorized on the right-hand side of Eq.~(\ref{cfac}).

The explicit expressions of ${\hat P}_{a_1 a_2}$,
for the splitting processes
\beq
\label{sppro}
a(p) \to a_1(zp + \kper + {\cal O}(\kper^2)) +
a_2((1-z) p - \kper + {\cal O}(\kper^2)) \;\;,
\eeq
depend on the flavour of the partons $a_1, a_2$ and are given 
by\footnote{The $\ep$ dependence on the right-hand side of 
Eqs.~(\ref{hpqqep})--(\ref{hpggep}) refers to the 
CDR 
scheme used throughout the paper. A detailed
discussion of the regularization-scheme dependence of the collinear splitting
functions at ${\cal O}(\as)$, including the corresponding explicit expressions,
can be found in Ref.~[\ref{schemedep}].}
\beeq
\label{hpqqep}
{\hat P}_{qg}^{s s'}(z,\kper;\ep) = {\hat P}_{{\bar q}g}^{s s'}(z,\kper;\ep)
= \delta_{ss'} \;C_F
\;\left[ \frac{1 + z^2}{1-z} - \ep (1-z) \right] \;\;,
\eeeq
\beeq
\label{hpqgep}
{\hat P}_{gq}^{s s'}(z,\kper;\ep) = {\hat P}_{g{\bar q}}^{s s'}(z,\kper;\ep)
= \delta_{ss'} \;C_F
\;\left[ \frac{1 + (1-z)^2}{z} - \ep z \right] \;\;,
\eeeq
\beeq
\label{hpgqep}
{\hat P}_{q{\bar q}}^{\mu \nu}(z,\kper;\ep) 
= {\hat P}_{{\bar q}q}^{\mu \nu}(z,\kper;\ep)
= T_R
\left[ - g^{\mu \nu} + 4 z(1-z) \frac{\kper^{\mu} \kper^{\nu}}{\kper^2}
\right] \;\;,
\eeeq
\beq
\label{hpggep}
{\hat P}_{gg}^{\mu \nu}(z,\kper;\ep) = 2C_A
\;\left[ - g^{\mu \nu} \left( \frac{z}{1-z} + \frac{1-z}{z} \right)
- 2 (1-\ep) z(1-z) \frac{\kper^{\mu} \kper^{\nu}}{\kper^2}
\right] \;\;,
\eeq
where the $SU(N_c)$ QCD colour factors are 
\beq
\label{colofac}
C_F = \frac{N_c^2 -1}{2N_c} \;, \;\;\; C_A = N_c \;, 
\;\;\; T_R = \frac{1}{2} \;,
\eeq
and the spin indices of the parent parton $a$ have been denoted by $s,s'$
if $a$ is a fermion and $\mu,\nu$ if $a$ is a gluon.

Note that when the parent parton is a fermion (see Eqs.~(\ref{hpqqep}) and
(\ref{hpqgep})) the splitting function is proportional to the unity matrix
in the spin indices. Thus, in the factorization formula (\ref{cfac}),
spin correlations are effective only in the case of the collinear splitting
of a gluon. Owing to the $\kper$-dependence of the gluon splitting
functions in Eqs.~(\ref{hpgqep}) and (\ref{hpggep}), these spin correlations
produce a non-trivial azimuthal dependence with respect to the directions
of the other momenta in the factorized matrix element.

Equations (\ref{hpqqep})--(\ref{hpggep}) lead to the more familiar form of the
$d$-dimensional splitting functions only after average over the polarizations
of the parton $a$. The $d$-dimensional average is obtained by means of the
factors in Eqs.~(\ref{ferav}) and (\ref{gluav}).
Denoting by $\la {\hat P}_{a_1 a_2} \ra$
the average of ${\hat P}_{a_1 a_2}$ over the polarizations of the parent 
parton $a$, we have:
\beeq
\label{avhpqq}
\la {\hat P}_{qg}(z;\ep) \ra \, = \, \la {\hat P}_{{\bar q}g}(z;\ep) \ra \,
= C_F\;\left[ \frac{1 + z^2}{1-z} - \ep (1-z) \right] \;\;,
\eeeq
\beeq
\label{avhpqg}
\la {\hat P}_{gq}(z;\ep) \ra \, = \, \la {\hat P}_{g{\bar q}}(z;\ep) \ra \,
= C_F \;\left[ \frac{1 + (1-z)^2}{z} - \ep z \right] \;\;,
\eeeq
\beeq
\label{avhpgq}
\la {\hat P}_{q{\bar q}}(z;\ep) \ra \, = \, \la {\hat P}_{{\bar q}q}(z;\ep) \ra
 \, = T_R \left[ 1 - \frac{2 z(1-z)}{1-\ep} \right] \;\;,
\eeeq
\beq
\label{avhpgg}
\la {\hat P}_{gg}(z;\ep) \ra \, = \, 2C_A
\;\left[ \frac{z}{1-z} + \frac{1-z}{z}
+ z(1-z) \right] \;\;.
\eeq

In the rest of this section we are mainly interested in the collinear behaviour
of the tree-level matrix element $\cm(p_1,\dots)$ in Eq.~(\ref{meldef})
at ${\cal O}(\as^2)$. At this order there are two different collinear
limits to be considered [\ref{glover}]. 

The first limit is approached when {\em two pairs}
of parton momenta, say $\{ p_1,p_2 \}$ and $\{ p_3,p_4 \}$, become parallel 
independently. In this case collinear factorization 
follows from the straightforward iteration of Eq.~(\ref{cfac}): the ensuing
factorization formula simply contains the product of the two splitting functions
${\hat P}_{a_1 a_2}^{s_{12}s_{12}^\prime}$ and 
${\hat P}_{a_3 a_4}^{s_{34}s_{34}^\prime}$.

In the second limit, three parton momenta can simultaneously become
parallel. This triple collinear limit is discussed in the following subsections.

\subsection{Kinematics in the triple collinear limit}
\label{kin}

We denote by $p_1, p_2$ and $p_3$ the momenta of the three collinear partons.
The most general parametrization of these collinear momenta is
\beq
\label{kin3}
p_i^\mu = x_i p^\mu +k_{\perp i}^\mu - \frac{k_{\perp i}^2}{x_i} 
\frac{n^\mu}{2p \cdot n} \;, \;\;\;\;\;i=1,2,3 \;,
\eeq
where, as in Eq.~(\ref{clim}), the light-like vector $p^\mu$ denotes 
the collinear direction and the auxiliary light-like vector $n^\mu$
specifies how the collinear direction is approached 
$(k_{\perp i} \cdot p = k_{\perp i} \cdot n = 0)$. 
Note that no other
constraint (in particular $\sum_i x_i \neq 1$ and $\sum_i k_{\perp i} \neq 0$) 
is imposed on the longitudinal and transverse variables $x_i$ and $k_{\perp i}$.
Thus, we can easily consider any (asymmetric) collinear limit at once.

Note, however, that the triple collinear limit is invariant 
under longitudinal boosts along the direction of the total momentum
$p_{123}^\mu = p_1^\mu + p_2^\mu + p_3^\mu$. Thus, the relevant kinematical
variables are the following boost-invariant quantities
\beeq
\label{zvar}
z_i &=& \frac{x_i}{\sum_{j=1}^3 \,x_j} \;\;,\\
\label{kvar}
{\ktil}_i^\mu &=& k_{\perp i}^\mu - \frac{x_i}{\sum_{k=1}^3 \,x_k} \;
\sum_{j=1}^3 k_{\perp j}^\mu \;\;.
\eeeq
Note that these variables automatically satisfy the constraints
$\sum_{i=1}^3 z_i = 1$ and  $\sum_{i=1}^3 {\ktil}_i = 0$, so that only four of
them are actually independent. 

In terms of the longitudinal and transverse variables introduced so far,
the two-particle sub-energies $s_{ij}$ are written as
\beq
\label{sijvar}
s_{ij}=(p_i+p_j)^2=- x_i x_j \left( \f{k_{\perp j}}{x_j}
-\f{k_{\perp i}}{x_i}\right)^2 
= - z_i z_j \left( \f{{\ktil}_j}{z_j} -\f{{\ktil}_i}{z_i}\right)^2 \;\;.
\eeq
It is also convenient to define the following variable $t_{ij,k}$
\beq
\label{tvar}
t_{ij,k} \equiv 2 \;\f{z_i s_{jk}-z_j s_{ik}}{z_i+z_j} +
\f{z_i-z_j}{z_i+z_j} \,s_{ij} \;\;.
\eeq

\subsection{Power counting and tree-level factorization at any order}
\label{power}

In the triple-collinear limit, the matrix element squared
$| \cm_{a_1,a_2,a_3,\dots}(p_1,p_2,p_3,\dots) |^2$ has the
singular behaviour
$| \cm_{a_1,a_2,a_3,\dots}(p_1,p_2,p_3,\dots) |^2 \sim 1/(s s')$, where
$s$ and $s'$ can be either two-particle $( s_{ij} = (p_i + p_j)^2 )$
or three-particle $( s_{123} = (p_1 + p_2 + p_3)^2 )$ sub-energies.
To define the collinear limit more precisely, we can rescale the transverse
momenta $k_{\perp i}$ by an overall factor $\lambda$:
\beq
\label{kscale}
k_{\perp i} \to \lambda \; k_{\perp i} \;, 
\eeq
and then perform the limit $\lambda \to 0$. In this limit the matrix element 
squared behaves as
\beq
\label{mscale}
| \cm_{a_1,a_2,a_3,\dots} |^2 \to {\cal O}(1/\lambda^4) + \dots \;,
\eeq
where the dots stand for less singular contributions when $\lambda \to 0$.
We are interested in explicitly evaluating the dominant singular term
${\cal O}(1/\lambda^4)$.

To study this singular behaviour we use power-counting
arguments and the universal factorization properties of collinear
singularities. The method [\ref{jetcalc}] is completely general: it is 
{\em process-independent} and can be applied to any {\em multiple} collinear 
limit $a \to a_1 \dots a_m$ at the tree level (i.e. at ${\cal O}(\as^{m-1})$).
Thus, we shall discuss the most general case with $m$ collinear partons.

We shall show that in the multiple collinear limit $a \to a_1 \dots a_m$,
the matrix element squared
$| \cm_{a_1,\dots,a_m,\dots}(p_1,\dots,p_m,\dots) |^2$
still fulfils a factorization formula analogous to Eq.~(\ref{cfac}), namely
\beeq
\label{ccfacm}
| \cm_{a_1,\dots,a_m,\dots}(p_1,\dots,p_m,\dots) |^2 \simeq
\left( \frac{8 \pi \mu^{2\ep} \as}{s_{1 \dots m}}\right)^{m-1}
 \;{\cal T}_{a,\dots}^{s s'}(xp,\dots) \;
{\hat P}_{a_1 \dots a_m}^{s s'}
\;\;,
\eeeq
where $s_{1 \dots m} = (p_1+\dots+p_m)^2$ is the $m$-particle sub-energy
and $x = \sum_{i=1}^m x_i$.

As in Eq.~(\ref{cfac}), the spin-polarization tensor 
${\cal T}_{a,\dots}^{s s'}(xp,\dots)$ is obtained by replacing the partons 
$a_1, \dots, a_m$ with a single parent parton, whose flavour $a$ is 
determined by flavour conservation in the splitting process.
More precisely, $a$ is a quark (antiquark) if the set $\{a_1, \dots, a_m\}$ 
contains an odd number of quarks (antiquarks), and $a$ is a gluon otherwise.

The factorization formula (\ref{ccfacm}) takes into account all the dominant
singular contributions in the multiple collinear limit, that is, all the
contributions that have the scaling behaviour $(1/\lambda^2)^{m-1}$ under the 
scale transformation in Eq.~(\ref{kscale}). Relative corrections of
${\cal O}(\lambda)$ are systematically neglected on the right-hand side
of Eq.~(\ref{ccfacm}).

The $m$-parton splitting functions ${\hat P}_{a_1 \dots a_3}$ are 
dimensionless functions of the parton momenta $p_1, \dots, p_m$ and
generalize the Altarelli--Parisi splitting functions in Eq.~(\ref{cfac}).
Owing to their invariance under longitudinal boosts along
the collinear direction, the splitting functions can depend in a non-trivial
way only on the sub-energy ratios $s_{ij}/s_{1 \dots m}$ and on the 
longitudinal- and transverse-momentum variables $z_i$ and ${\ktil}_i$ defined
by the generalization of Eqs.~(\ref{zvar}) and~(\ref{kvar}) to the $m$-parton 
case.

The spin 
correlations produced by the collinear splitting are taken into account by
the splitting functions in a universal way, i.e. independently of the 
specific matrix element on the right-hand side of Eq.~(\ref{ccfacm}).

In the case of the splitting processes that involve a fermion as parent parton,
we find that spin correlations are absent. We can thus write the 
corresponding spin-dependent splitting function in terms of its average
$\la \Ph_{a_1 \dots a_m} \ra$ over the polarizations of the parent fermion $a$:
\beq
\label{qmaver}
\Ph^{ss'}_{a_1 \dots a_m} = \delta^{ss'} \,
\la \Ph_{a_1 \dots a_m} \ra \;\;.
\eeq
This feature is completely analogous to the 
${\cal O}(\as)$ case and follows from helicity conservation in 
the quark--gluon vector coupling.

In the case of collinear decays of a parent gluon, however,
spin correlations are highly non-trivial. 

Note also that the splitting functions for the collinear decay of an antiquark
can be simply obtained by charge-conjugation invariance from those of the
corresponding quark decay process, i.e. 
$\la \Ph_{a_1 \dots a_m} \ra = \la \Ph_{{\bar a}_1 \dots {\bar a}_m} \ra$.

\begin{figure}[htb]
\begin{center}
\begin{tabular}{c}
\epsfxsize=9truecm
\epsffile{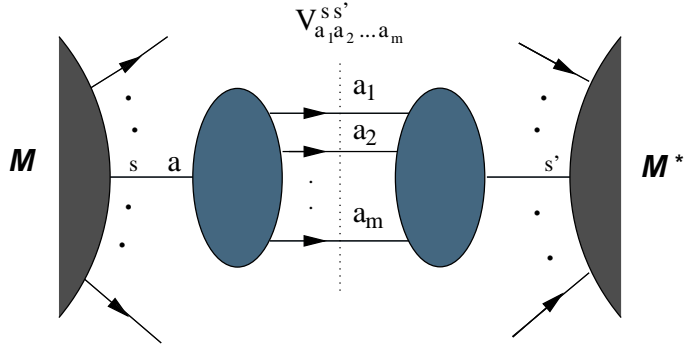}\\
\end{tabular}
\end{center}
\caption{\label{relevant}{\em Dominant diagrams for the multiple collinear
    limit at ${\cal O}(\as^{m-1})$ in a physical gauge.}}
\end{figure}

\begin{figure}[htb]
\begin{center}
\begin{tabular}{c}
\epsfxsize=9truecm
\epsffile{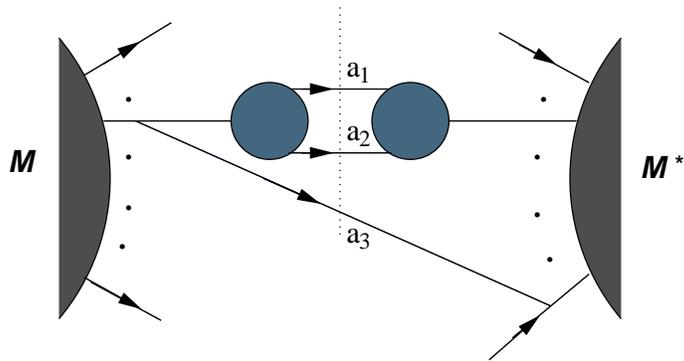}\\
\end{tabular}
\end{center}
\caption{\label{irrelevant}{\em An interference diagram for the triple 
collinear limit $a\to a_1a_2a_3$.
    }}
\end{figure}

The method used to derive these results
exploits the basic observation [\ref{collpc}] that
interfering Feynman diagrams obtained by squaring the amplitude 
$\cm(p_1,\dots,p_m,\dots)$ are collinearly suppressed when computed
in a physical gauge. Thus, in the evaluation of the multiple
collinear limit we can write
\beeq
| \cm_{a_1,\dots,a_m,\dots}(p_1,\dots,p_m,\dots) |^2 &\simeq&
\left[ \cm_{a,\dots}^{(n)}(p_1+\dots+p_m,\dots)
\right]^{\dagger} \;{\cal V}_{a_1\dots a_m}^{(n)}(p_1,\dots,p_m) \nonumber \\
\label{fcollgen}
&\cdot& \cm_{a,\dots}^{(n)}(p_1+\dots+p_m,\dots) + \dots \;\;.
\eeeq
The first term on the right-hand side of Eq.~(\ref{fcollgen}) corresponds
to the non-interfering Feynman diagrams in Fig.~\ref{relevant},
while the dots stand for subdominant contributions coming from interferences
(see e.g. the diagram in Fig.~\ref{irrelevant}). The superscripts $(n)$
denote that the various terms are evaluated in a physical gauge.
To simplify the calculation it is convenient to choose the axial gauge
$n\cdot A = 0$, where the gauge vector $n^\mu$ coincides with the auxiliary
light-like vector used in Eq.~(\ref{kin3}) to parametrize the collinear 
kinematics. The corresponding gluon polarization tensor $d^{\mu \nu}_{(n)}$
is
\beq
\label{dgauge}
d^{\mu \nu}_{(n)}(q) = - g^{\mu \nu} 
+ \frac{n^\mu q^\nu + q^\mu n^\nu}{n\cdot q} \;\;,
\eeq
where $q$ is the gluon momentum.

The summation over spin and colour indices is understood on
the right-hand side of Eq.~(\ref{fcollgen}).
The function ${\cal V}_{a_1 \dots a_m}$ in Eq.~(\ref{fcollgen}) is the
$m$-parton dispersive contribution to the two-point function of
the parent parton $a$. Being a two-point function, it is proportional to
the unity matrix in the colour indices of the parton $a$. Thus, we can sum
over the colours of the partons in the tree-level amplitudes, and we can
rewrite Eq.~(\ref{fcollgen}) in terms of the spin-polarization tensor
${\cal T}_{a,\dots}$ introduced in Eq.~(\ref{melspindef}):
\beq
\label{fcollimp}
| \cm_{a_1,\dots,a_m,\dots}(p_1,\dots,p_m,\dots) |^2 \simeq
\left( \frac{8 \pi \mu^{2\ep} \as}{s_{1 \dots m}}\right)^{m-1} 
\;{\cal T}_{a,\dots}^{(n)}(p_1+\dots+p_m,\dots) \;
V_{a_1 \dots a_m}^{(n)}(p_1,\dots,p_m)
\;.
\eeq
The function $V_{a_1 \dots a_m}^{(n)}$ is simply obtained from 
${\cal V}_{a_1 \dots a_m}^{(n)}$ in Eq.~(\ref{fcollgen}) by performing the
average over the colours of the parent parton $a$ and extracting the factor
in the round bracket on the right-hand side of Eq.~(\ref{fcollimp}).
Thus the tree-level function 
$V_{a_1 \dots a_m}^{(n)}(p_1,\dots,p_m)$ does not contain
any additional power of the QCD coupling $\as$. Note also that the spin tensor
${\cal T}_{a,\dots}^{(n)}(p_1+\dots+p_m,\dots)$ is not yet exactly
the physical polarization tensor of Eq.~(\ref{ccfacm}). In fact,
the momentum of the parton $a$ is off-shell ($(p_1+\dots+p_m)^2 = s_{1 \dots m}
\neq 0$) and, thus,
 ${\cal T}_{a,\dots}^{(n)}(p_1+\dots+p_m,\dots)$ is gauge-dependent.

To proceed, we should consider separately the two cases in which
the parton $a$ is either a quark (or antiquark) or a gluon.

\bigskip
\noindent {\it Quark splitting processes}

\noindent It is convenient to include the spin-polarization matrices
${\slash p}_1+ \dots+ {\slash p}_m$ of the decaying quark $a=q$ in the 
definition of the Dirac matrix $V_{a_1 \dots a_m}^{(n)}(p_1,\dots,p_m)$.
The most general decomposition of $V_{a_1 \dots a_m}^{(n)}(p_1,\dots,p_m)$ is
\beq
\label{aqdec}
V_{a_1 \dots a_m}^{(n)}(p_1,\dots,p_m) \sim \sum 
\Bigl( {\rm scalar \; amplitude} \Bigr) \cdot 
\Bigl( {\rm string \;of \;gamma \;matrices} \Bigr) \;\;.
\eeq
Any string of gamma matrices is obtained by multiplying an arbitrary
number of terms ${\slash v}_l$ with $l=1, \dots,m+1$, where ${\slash v}$ can 
be either ${\slash v}_i={\slash p}_i$ or 
${\slash v}_{m+1} = {\slash n}s_{1 \dots m}/n \cdot(p_1+\dots+p_m)$.
The matrices ${\slash v}_l$, like 
$V_{a_1 \dots a_m}^{(n)}(p_1,\dots,p_m)$, are homogeneous functions of $n^\mu$
with vanishing homogeneity degree. Thus, by Lorentz covariance,
the amplitudes on the right-hand side
of Eq.~(\ref{aqdec}) are scalar functions of the sub-energies $s_{ij}$ and the
longitudinal-momentum fractions $z_i= n \cdot p_i /n \cdot(p_1+\dots+p_m)$.
Moreover, they are rational functions of the variables $s_{ij}, z_i$ and thus,
by dimensional analysis, the corresponding strings 
can contain only an {\em odd} number of gamma matrices. 

We can now exploit the hermiticity properties of 
$V_{a_1 \dots a_m}^{(n)}(p_1,\dots,p_m)$. Since the scalar amplitudes are real,
the strings of gamma matrices appear in the form 
\beq
\label{strings}
\left( \frac{1}{s_{1 \dots m}} \right)^{(k-1)/2}
\left[ \;
{\slash v}_{i_1} {\slash v}_{i_2}\cdots {\slash v}_{i_k} +  
{\slash v}_{i_k} \cdots {\slash v}_{i_2} {\slash v}_{i_1} \; \right] \;\;,
\eeq
where the normalization by the overall power of $1/s_{1 \dots m}$ has been 
introduced to make the scalar amplitudes on the right-hand side
of Eq.~(\ref{aqdec}) dimensionless. Owing to the fact that
$k$ is odd, the terms with $k=3,7,11, \ldots$ in Eq.~(\ref{strings})
can in turn be reduced to strings that contain $k=1,5,9, \ldots$ gamma matrices
by using the anticommuting properties of the Dirac algebra.
This is the simplest form in which we can write the general decomposition of
Eq.~(\ref{aqdec}). 

We can now discuss separately the cases that involve the collinear decay 
of less or more than four partons.

From the previous discussion we conclude that, when $m \leq 3$, the 
functions $V_{a_1 \dots a_m}^{(n)}$
can be written as follows
\beq
\label{aqdecfin}
V_{a_1 \dots a_m}^{(n)}(p_1,\dots,p_m) = \sum_{i=1}^m 
A_i^{(q)}(\{s_{jl}, z_j\}) \;{\slash p}_i + B^{(q)}(\{s_{jl}, z_j\}) 
\;\frac{{\slash n} \;s_{1 \dots m}}{n \cdot(p_1+\dots+p_m)} \;, \;\;(m \leq 3)
\, ,
\eeq
while, when $m= 4$, we have
\beeq
\label{aqdecfin4}
V_{a_1 \dots a_4}^{(n)}(p_1,\dots,p_4) &=& \sum_{i=1}^4 
A_i^{(q)}(\{s_{jl}, z_j\}) \;{\slash p}_i + B^{(q)}(\{s_{jl}, z_j\}) 
\;\frac{{\slash n} \;s_{1 \dots 4}}{n \cdot(p_1+\dots+p_4)} \\
&+& C^{(q)}(\{s_{jl}, z_j\}) 
\;\frac{{\slash p}_1 {\slash p}_2 {\slash p}_3 {\slash p}_4 {\slash n} 
+ {\slash n} {\slash p}_4 {\slash p}_3 {\slash p}_2 {\slash p}_1}{s_{1 \dots 4}
\; n \cdot(p_1+\dots+p_4)} \;. 
\eeeq
Then we can proceed to single out the dominant singular behaviour of
Eqs.~(\ref{aqdecfin}) and (\ref{aqdecfin4}) in the multiple collinear limit.
Since the scalar functions $A_i^{(q)}(\{s_{jl}, z_j\}), 
B^{(q)}(\{s_{jl}, z_j\})$ and 
$C^{(q)}(\{s_{jl}, z_j\})$ are dimensionless, they are
invariant under the scale transformation (\ref{kscale}). Moreover,
since we can write
\beq
\label{piki}
p_i^\mu=z_i (p_1+ \dots +p_m)^\mu+{\ktil}_i^\mu+{\cal O}(k_{\perp}^2) \;\;,
\eeq
by rescaling
the transverse momenta as in Eq.~(\ref{kscale}) we obtain the
following scaling behaviour  
\beq
\label{aqscale}
V_{a_1 \dots a_m}^{(n)}(p_1,\dots,p_m) =  ( {\slash p}_1 +\dots + {\slash p}_m )
\; \sum_{i=1}^m  z_i \, A_i^{(q)}(\{s_{jl}, z_j\}) 
\left[ 1 + {\cal O}(\lambda) \right]
\;, \;\;\;(m \leq 4) \,. 
\eeq
Thus, inserting Eq.~(\ref{aqscale}) into Eq.~(\ref{fcollimp}), we can use the 
spin polarization factor ${\slash p}_1 +\dots + {\slash p}_m$ to reconstruct 
the matrix element squared $| \cm_{q,\dots}^{(n)}(p_1+\dots+p_m,\dots) |^2$. 
Having already factorized the dominant singular term, we can now replace
$p_1+\dots+p_m \to xp$ in $| \cm_{q,\dots}^{(n)} |^2$, so that its gauge
dependence disappears, and we obtain the factorization formula (\ref{ccfacm}).
Moreover, we also obtain an explicit expression for the quark splitting function
in terms of the scalar amplitudes $A_i^{(q)}$ in Eqs.~(\ref{aqdecfin}) and 
(\ref{aqdecfin4}):
\beq
\label{qmaver4}
\Ph^{ss'}_{a_1 \dots a_m} = \delta^{ss'} \,
\sum_{i=1}^m  z_i \, A_i^{(q)}(\{s_{jl}, z_j\}) \;, \;\;\;\; (m \leq 4) \;.
\eeq 

This argument to prove collinear factorization is based on the fact that a
single spin structure (see Eq.~(\ref{aqscale})) dominates the collinear limit 
of the quark decay function $V_{a_1 \dots a_m}^{(n)}$. In particular, this
implies that spin correlations are
absent from the collinear decay of a fermion, independently of the number of 
its spin polarizations. 
However, the argument works only for the cases with $m \leq 4$. 
When $m > 4$ collinear factorization still applies but, as shown below, 
spin correlations cancel only if we use a dimensional-regularization
prescription in which the massless fermion has two spin polarizations.

According to our definition, the scalar amplitudes on the right-hand side of 
Eq.~(\ref{aqdec}) are dimensionless and, hence, they are invariant 
under the scale transformation (\ref{kscale}). The collinear limit of 
Eq.~(\ref{aqdec}) is thus determined by that of the strings of gamma matrices 
in Eq.~(\ref{strings}). Using Eq.~(\ref{piki}) and rescaling
the transverse momenta as in Eq.~(\ref{kscale}), we see that the strings that
dominate in the multiple collinear limit are those of the form
\beeq
&&\left( \frac{1}{s_{1 \dots m}} \right)^{k} \,\left[
\;({\slash p}_1 + \dots {\slash p}_m) \, {\slash {\ktil}}_{i_1}
\, {\slash {\ktil}}_{i_2} \cdots {\slash {\ktil}}_{i_{2k}}
+ {\slash {\ktil}}_{i_{2k}} \cdots {\slash {\ktil}}_{i_2} \,
{\slash {\ktil}}_{i_1} \,({\slash p}_1 + \dots {\slash p}_m) \right] \nn \\
\label{stringscoll}
&&= \left( \frac{1}{s_{1 \dots m}} \right)^{k} \; x 
\,\left[ {\slash p}
\, {\slash {\ktil}}_{i_1} \, {\slash {\ktil}}_{i_2} \cdots 
{\slash {\ktil}}_{i_{2k}} +
{\slash {\ktil}}_{i_{2k}} \cdots {\slash {\ktil}}_{i_2} \,
{\slash {\ktil}}_{i_1} \,{\slash p} \right]
\;\left[ 1 + {\cal O}(\lambda) \right] \;\;,
\eeeq
where the dots stand for the product of ${\slash {\ktil}}_i$ factors. We can
now multiply Eq.~(\ref{stringscoll}) by unity in the form 
$1 =  ({\slash p} {\slash n} + {\slash n}{\slash p})/(2p\cdot n)$, and,
using $\{ {\slash p} , {\slash {\ktil}}_i \} =0$ and
${\slash p}^2 =0$,
we obtain
\beeq
&&\left( \frac{1}{s_{1 \dots m}} \right)^{k} \; x \left[ {\slash p}
\, {\slash {\ktil}}_{i_1} \, {\slash {\ktil}}_{i_2} \cdots 
{\slash {\ktil}}_{i_{2k}} 
\;\frac{{\slash p} {\slash n} + {\slash n} {\slash p}}{2p\cdot n}
+ \frac{{\slash p} {\slash n} + {\slash n} {\slash p}}{2p\cdot n}
{\slash {\ktil}}_{i_{2k}} \cdots {\slash {\ktil}}_{i_2} \,
{\slash {\ktil}}_{i_1} \,{\slash p} \right] \nn \\
\label{stringdom}
&&= \left( \frac{1}{s_{1 \dots m}} \right)^{k} \; x {\slash p}
\; \frac{{\slash {\ktil}}_{i_1} \, {\slash {\ktil}}_{i_2} \cdots 
{\slash {\ktil}}_{i_{2k}} \,{\slash n} +
{\slash n} \,{\slash {\ktil}}_{i_{2k}} \cdots {\slash {\ktil}}_{i_2} \,
{\slash {\ktil}}_{i_1} }{2xp\cdot n} \; x{\slash p} \;\;.
\eeeq
Denoting by $\chi_s(p)$ the spinor of an on-shell fermion with momentum
$p$ and spin polarization $s$, we then replace the polarization matrices
$x{\slash p}$ in Eq.~(\ref{stringdom}) by using the identity
$x{\slash p} = \sum_s \chi_s(xp) {\overline \chi}_s(xp)$ and we can rewrite 
the string in Eq.~(\ref{stringdom}) as follows
\beq
\label{stringspin}
\sum_{s,s^\prime} \left[ \;\chi_s(xp) \;{\overline \chi}_{s^\prime}(xp) 
\;\right]
\; \left( \frac{1}{s_{1 \dots m}} \right)^{k}
{\overline \chi}_s(xp) \;\frac{{\slash {\ktil}}_{i_1} \, {\slash {\ktil}}_{i_2} 
\cdots {\slash {\ktil}}_{i_{2k}} \,{\slash n} + {\slash n} \,
{\slash {\ktil}}_{i_{2k}} \cdots {\slash {\ktil}}_{i_2} \, 
{\slash {\ktil}}_{i_1}}{2xp\cdot n} \;\chi_{s^\prime}(xp)
\;\;.
\eeq
When inserted in Eqs.~(\ref{aqdec}) and (\ref{fcollimp}),
the factor in the square bracket reconstructs the polarization matrix of the
decaying quark and, thus, the spin-polarization tensor
${\cal T}_{q,\dots}^{s s^\prime}(xp,\dots)$ in the factorization formula
(\ref{ccfacm}). The remaining factor in Eq.~(\ref{stringspin}) gives the
contribution of the string in Eq.~(\ref{stringscoll}) to the
quark splitting function ${\hat P}^{s s^\prime}_{a_1 \dots a_m}$.

By explicit construction we see that in general the splitting function
${\hat P}^{s s^\prime}_{a_1 \dots a_m}$ is not diagonal with respect to the
spin indices. Nonetheless, the spin correlations are absent within the 
dimensional-regularization prescription used throughout the paper. Since we
are considering only two helicity states for massless quarks, we have
$\chi_{s=\pm 1}(p) = \frac{1}{2}(1 \pm \gamma_5) \chi(p)$, 
where $\chi(p)$ is a generic Dirac spinor. Thus, using the general properties 
of the Dirac algebra, the contribution of Eq.~(\ref{stringspin}) to the
splitting function can straightforwardly be recast in a form that explicitly
shows the cancellation of the spin correlations:
\beeq
&&\left( \frac{1}{s_{1 \dots m}} \right)^{k}
{\overline \chi}_s(xp) \;\frac{{\slash {\ktil}}_{i_1} \, {\slash {\ktil}}_{i_2} 
\cdots {\slash {\ktil}}_{i_{2k}} \,{\slash n} + {\slash n} \,
{\slash {\ktil}}_{i_{2k}} \cdots {\slash {\ktil}}_{i_2} \, 
{\slash {\ktil}}_{i_1}}{2xp\cdot n} \;\chi_{s^\prime}(xp) \nn \\
\label{stringhel}
&&= \delta^{s s^\prime} \left( \frac{1}{s_{1 \dots m}} \right)^{k}
{\rm Tr} \left[ \frac{{\slash n} \left( {\slash p} \, {\slash {\ktil}}_{i_1} 
\, {\slash {\ktil}}_{i_2} \cdots {\slash {\ktil}}_{i_{2k}} +
{\slash {\ktil}}_{i_{2k}} \cdots {\slash {\ktil}}_{i_2} \,
{\slash {\ktil}}_{i_1} \, {\slash p} \right)}{4p\cdot n} \right]
\;\;, \;\;\;s,s^\prime=\pm 1 \;\;,
\eeeq
where ${\rm Tr}$ denotes the trace of the Dirac matrices.
The identity in Eq.~(\ref{stringhel}) relies on the definition and the
properties of the chiral projectors $\frac{1}{2}(1 \pm \gamma_5)$ and
the absence of spin correlations ultimately follows from helicity conservation 
in the quark--gluon vector coupling.

This method to derive collinear factorization also provides us with
an expression of the (spin-averaged)
{\em quark} splitting function in terms of the dispersive 
part $V_{a_1 \dots a_m}^{(n)}(p_1,\dots,p_m)$ of the two-point quark amplitude.
From Eqs.~(\ref{aqdec}) and (\ref{stringspin}), we find
\beq
\label{pqvsa}
\la \Ph_{a_1 \dots a_m} \ra  
= {\rm Tr} \left( \frac{{\slash n} \;
V_{a_1 \dots a_m}^{(n)}(p_1,\dots,p_m)}{4 \,n \cdot(p_1+\dots+p_m) } 
\right) \;\;.
\eeq
This equation is useful for a straightforward evaluation of the splitting
function for the multiple collinear decay of a quark.

\bigskip
\noindent {\it Gluon splitting processes}

\noindent Unlike the quark case, it is convenient to define the gluon
two-point function $V_{a_1 \dots a_m}^{(n)}$
without including in it the spin-polarization tensors 
$d^{\mu \nu}_{(n)}(p_1+\dots+p_m)$ of the two external gluons.
Because of Lorentz covariance and the vanishing degree of homogeneity with
respect to $n^\mu$, the spin tensor $V_{a_1 \dots a_m}^{\nu \rho \, (n)}$
can be decomposed in terms of dimensionless scalar amplitudes as 
\beeq
&&\!\! \!\! \! \!\!\! \!\! \!\!\! \!\!
V_{a_1 \dots a_m}^{\mu \nu \,(n)}(p_1,\dots,p_m) = 
A^{(g)}(\{s_{jl}, z_j\})
\; g^{\mu \nu} + \sum_{i,j=1}^m B_{i,j}^{(g)}(\{s_{kl}, z_k\}) \;
\frac{p_i^\mu p_j^\nu}{s_{1 \dots m}} \nn \\
\label{agdec}
&\!\!+& \sum_{i=1}^m C_{i}^{(g)}(\{s_{jl}, z_j\}) \;
\frac{p_i^\mu n^\nu + n^\mu p_i^\nu}{n \cdot(p_1+\dots+p_m)}
+ D^{(g)}(\{s_{jl}, z_j\}) \;
\frac{n^\mu n^\nu \, s_{1 \dots m}}{(n \cdot(p_1+\dots+p_m))^2}
\;.
\eeeq
Then, we have to multiply $V_{a_1 \dots a_m}^{\nu \rho \, (n)}$
by the gluon polarization tensors as follows
\beq
\label{agdecfin}
d^{\mu}_{\nu \,(n)}(p_1+\dots+p_m) 
\;V_{a_1 \dots a_m}^{\nu \rho \,(n)}(p_1,\dots,p_m)
\; d^{\sigma}_{\rho \,(n)}(p_1+\dots+p_m) \;\;.
\eeq
Inserting Eq.~(\ref{agdec}) into Eq.~(\ref{agdecfin}), we immediately see
that the scalar amplitudes $C_{i}^{(g)}(\{s_{jl}, z_j\})$ and 
$D^{(g)}(\{s_{jl}, z_j\})$ give a vanishing contribution because the gauge 
vector $n^\mu$ is orthogonal to the polarization tensors. 
As for the second term on the right-hand side of Eq.~(\ref{agdec}), we can
extract its dominant collinear contribution by simply performing 
the replacement $p_i^\mu \to {\ktil}_i^\mu$. Indeed, using Eq.~(\ref{piki})
and 
\beq
(p_1+\dots+p_m)_\mu d_{(n)}^{\mu\nu}(p_1+\dots+p_m)={\cal O}(s_{1 \dots m})
\;\;,
\eeq
we have
\beq
d^{\mu}_{\nu \,(n)}(p_1+\dots+p_m) \; p_i^\nu \; p_j^\rho
\; d^{\sigma}_{\rho \,(n)}(p_1+\dots+p_m) = {\ktil}_i^\mu \;{\ktil}_i^\sigma
+{\cal O}(\lambda^3) \;\;,
\eeq
so that the longitudinal component of $p_i^\mu$ is suppressed in the
multiple collinear limit $\lambda \to 0$.

We can now safely perform the collinear limit and we obtain 
the factorization formula (\ref{ccfacm}) and an explicit expression
of the gluon splitting function in terms of the scalar amplitudes in 
Eq.~(\ref{agdec}):
\beq
\label{pggen}
\Ph^{\mu \nu}_{a_1 \dots a_m} =
A^{(g)}(\{s_{jl}, z_j\})
\; g^{\mu \nu} + \sum_{i,j=1}^m B_{i,j}^{(g)}(\{s_{kl}, z_k\}) \;
\frac{{\ktil}_i^\mu {\ktil}_j^\nu}{s_{1 \dots m}} \;\;.
\eeq

The splitting function 
can be averaged over the spin polarizations of the parent
gluon according to Eq.~(\ref{gluav}), and we obtain
\beq
\label{gsfav}
\la \Ph_{a_1 \dots a_m} \ra \, \equiv \frac{1}{2 (1 - \ep)} \,d_{\mu \nu}(p)
\; \Ph^{\mu\nu}_{a_1 \dots a_m} 
= - A^{(g)}(\{s_{jl}, z_j\}) - \, \frac{1}{2 (1 - \ep)}
\sum_{i,j=1}^m B_{i,j}^{(g)}(\{s_{kl}, z_k\}) \;
\frac{{\ktil}_i \cdot {\ktil}_j}{s_{1 \dots m}}  \;,
\eeq
where
\beq
2 {\ktil}_i \cdot {\ktil}_j = s_{ij} 
- \sum_{k=1}^m \left( z_i s_{jk} + z_j s_{ik} \right) + 2 z_i z_j s_{1 \dots m}
\;\;.
\eeq
Note that, since $2 {\ktil}_i \cdot {\ktil}_j =
2 p_i \cdot  d_{(n)}(p_1+\dots+p_m) \cdot d_{(n)}(p_1+\dots+p_m) \cdot p_j$,
the spin-averaged splitting function can also be expressed in terms
of the Lorentz trace of Eq.~(\ref{agdecfin}):
\beq
\la \Ph_{a_1 \dots a_m} \ra \, = - \frac{1}{2 (1 - \ep)} \;
d^{\mu}_{\nu \,(n)}(p_1+\dots+p_m) 
\;V_{a_1 \dots a_m}^{\nu \rho \,(n)}(p_1,\dots,p_m)
\; d_{\rho \mu \,(n)}(p_1+\dots+p_m) \;\;.
\eeq

In the following subsection we present the explicit calculation of
the quark and gluon splitting functions in the triple collinear limit.

\subsection{Collinear splitting functions at ${\cal O}(\as^2)$}
\label{splitt}

The list of (non-vanishing) splitting processes that we have to consider is
as follows:
\beeq
\label{qqqprime}
&& q\to {\bar q}^\prime_1 + q^\prime_2 + q_3 \;\;, 
\;\;({\bar q} \to {\bar q}^\prime_1 + q^\prime_2 + {\bar q}_3 ) \;\;, \\
\label{qqq}
&& q\to {\bar q}_1 + q_2 + q_3 \;\;, 
\;\;({\bar q} \to {\bar q}_1 + q_2 + {\bar q}_3 )\;\;, \\
\label{ggq}
&& q\to g_1 + g_2 + q_3 \;\;,
\;\;({\bar q} \to g_1 + g_2 + {\bar q}_3 ) \;\;, \\
\label{gqq}
&& g\to g_1 + q_2 + {\bar q}_3 \;\;, \\
\label{ggg}
&& g \to g_1 + g_2 + g_3 \;\;. 
\eeeq
The superscripts in $q^\prime, {\bar q}^\prime$ denote fermions with different
flavour with respect to $q, {\bar q}$. As already mentioned in
Sect.~\ref{power}, the splitting functions for the
processes in parenthesis in Eqs.~(\ref{qqqprime}) and (\ref{qqq}) can be simply
obtained by charge-conjugation invariance, i.e. 
$\Ph_{{\bar q}^\prime_1 q^\prime_2 {\bar q}_3} =
\Ph_{q^\prime_1 {\bar q}^\prime_2 q_3}$ and
$\Ph_{{\bar q}_1 q_2 {\bar q}_3} =
\Ph_{q_1 {\bar q_2} q_3}$.
In summary, we have to compute five independent splitting functions.

\begin{figure}[htb]
\begin{center}
\begin{tabular}{c}
\epsfxsize=5truecm
\epsffile{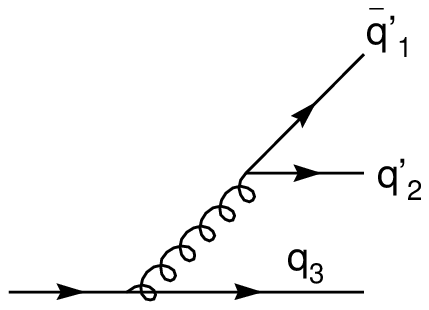}\\
\end{tabular}
\end{center}
\caption{\label{qqqnid}{\em The diagram for the collinear decay
$q\to {\bar q^\prime}_1q^\prime_2q_3$. }}
\end{figure}

To illustrate our calculation, 
we first consider the process in Eq.~(\ref{qqqprime}), that is, 
the case in  which a quark--antiquark pair ${\bar q^\prime}_1q^\prime_2$ 
and a quark $q_3$ with different flavour become collinear.
This is the simplest case, because the two-point function
$V_{{\bar q}^\prime_1 q^\prime_2 q_3}^{(n)}(p_1,p_2,p_3)$
for the corresponding splitting process is obtained by 
squaring the sole Feynman diagram in Fig.~\ref{qqqnid}.
According to the definition in Eqs.~(\ref{fcollimp}) and (\ref{aqdec}), we
extract the overall factor  $\left( 8 \pi \mu^{2\ep} \as/s_{123} \right)^2$
and, performing the average over the colours of the decaying quark $q$,
we find
\beeq
V_{{\bar q}^\prime_1 q^\prime_2 q_3}^{(n)}(p_1,p_2,p_3)
&=& \f{1}{2} \, 
C_F T_R \,\f{s_{123}}{s_{12}} \left\{ \left[ \f{2z_3}{z_1+z_2}
- \left( \f{t_{12,3}^2}{s_{12}^2} + 1 - 2\ep \right) \f{s_{12}}{s_{123}}
\right] ( {\slash p}_1 + {\slash p}_2 + {\slash p}_3 ) \right. \nn \\
\label{prelimit}
&+& \left. \f{2}{z_1+z_2}{\slash p}_3 + (1-2\ep) ({\slash p}_1 + {\slash p}_2)
+ \f{z_1 - z_2}{z_1+z_2} ({\slash p}_1 - {\slash p}_2) \right. \\
&+& \left. \f{2t_{12,3}}{(z_1+z_2)s_{12}} (z_1{\slash p}_2 - z_2{\slash p}_1) 
+ \f{1}{z_1+z_2} \left( \f{s_{12}}{s_{123}} -1 \right) 
\f{ {\slash n} \;s_{123}}{n\cdot(p_1+p_2+p_3)} \right\} \nn
\;\;,
\eeeq
where $t_{12,3}$ is the kinematical variable defined in Eq.~(\ref{tvar}).
Note that Eq.~(\ref{prelimit}) has the general structure obtained in
Eq.~(\ref{aqdecfin}). Using Eq.~(\ref{pqvsa}) to compute the splitting function
$\la \Ph_{{\bar q}^\prime_1 q^\prime_2 q_3} \ra$,
the last two terms on the right-hand side of
Eq.~(\ref{prelimit}) give a vanishing contribution and we obtain
the final result:
\beq
\label{qqqprimesf}
\la \Ph_{{\bar q}^\prime_1 q^\prime_2 q_3} \ra \, = \f{1}{2} \, 
C_F T_R \,\f{s_{123}}{s_{12}} \left[ - \f{t_{12,3}^2}{s_{12}s_{123}}
+\f{4z_3+(z_1-z_2)^2}{z_1+z_2} 
+ (1-2\ep) \left(z_1+z_2-\f{s_{12}}{s_{123}}\right)
\right] \;\;.
\eeq

The calculation of the splitting functions for the other processes
in Eqs.~(\ref{qqq})--(\ref{ggg}) can be performed exactly in the same manner,
by using the general procedure discussed in Sect.~\ref{power}.
We first compute the corresponding two-point functions
$V_{a_1 a_2 a_3}^{(n)}(p_1,p_2,p_3)$ and then, using Eqs.~(\ref{pqvsa}) and
(\ref{pggen}), we evaluate the splitting functions. Since the intermediate
expressions for the two-point functions are quite cumbersome, in the following
we limit ourselves to showing the relevant Feynman diagrams and to presenting
the final results for the splitting functions.

\begin{figure}[htb]
\begin{center}
\begin{tabular}{c}
\epsfxsize=8.5truecm
\epsffile{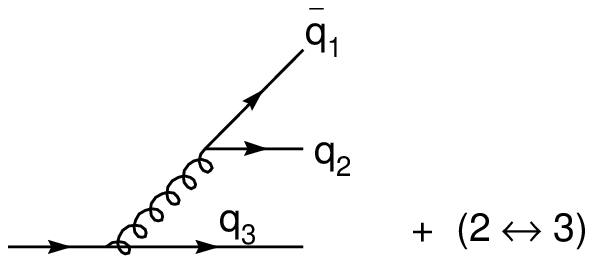}\\
\end{tabular}
\end{center}
\caption{\label{qqqid}{\em The diagrams for the collinear decay
$q\to {\bar q}_1q_2q_3$. }}
\end{figure}

The calculation of the splitting function for the case of final-state fermions 
with identical flavour involves a diagram analogous to that in 
Fig.~\ref{qqqnid} plus its crossed diagram (see Fig.~\ref{qqqid}). Thus,
the result can be written in terms of that in Eq.~(\ref{qqqprimesf}),
as follows
\beq
\label{qqqsf}
\la \Ph_{{\bar q}_1q_2q_3} \ra \, =
\left[ \la \Ph_{{\bar q}^\prime_1q^\prime_2q_3} \ra \, + \,(2\lra 3) \,\right]
+ \la \Ph^{({\rm id})}_{{\bar q}_1q_2q_3} \ra \;\;,
\eeq 

where the interference contribution is given by
\beeq
\label{idensf}
\la \Ph^{({\rm id})}_{{\bar q}_1q_2q_3} \ra \,
&=& C_F \left( C_F-\f{1}{2} C_A \right)
 \Biggl\{ (1-\ep)\left( \f{2s_{23}}{s_{12}} - \ep \right)\nn\\
&+& \f{s_{123}}{s_{12}}\Biggl[\f{1+z_1^2}{1-z_2}-\f{2z_2}{1-z_3}
    -\ep\left(\f{(1-z_3)^2}{1-z_2}+1+z_1-\f{2z_2}{1-z_3}\right) 
- \ep^2(1-z_3)\Biggr] \nn\\
&-& \f{s_{123}^2}{s_{12}s_{13}}\f{z_1}{2}\left[\f{1+z_1^2}{(1-z_2)(1-z_3)}-\ep
    \left(1+2\f{1-z_2}{1-z_3}\right)
    -\ep^2\right] \Biggr\} + (2\lra 3) \;\;.
\eeeq

\begin{figure}[htb]
\begin{center}
\begin{tabular}{c}
\epsfxsize=13truecm
\epsffile{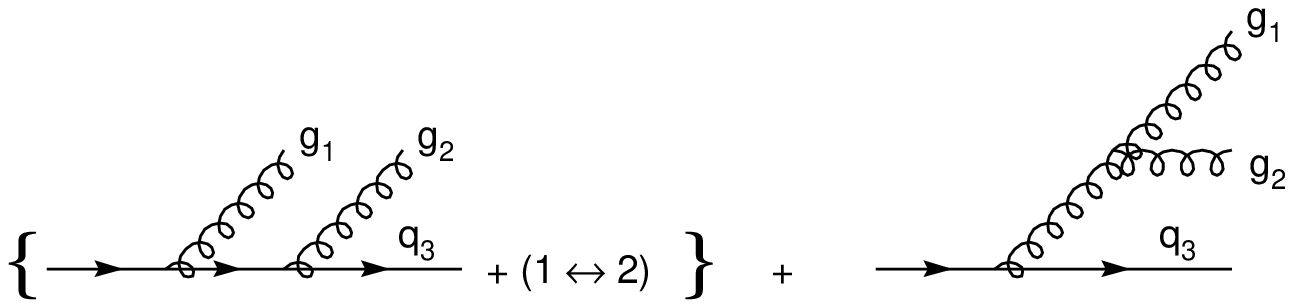}\\
\end{tabular}
\end{center}
\caption{\label{qgg}{\em The diagrams for the collinear decay
$q\to g_1g_2q_3$. }}
\end{figure}

The splitting function of the remaining quark-decay subprocess is obtained
by squaring the diagrams in Fig.~\ref{qgg}. It can be
decomposed according to the different colour coefficients:
\beq
\label{qggsf}
\la \Ph_{g_1 g_2 q_3} \ra \, =
C_F^2 \, \la \Ph_{g_1 g_2 q_3}^{({\rm ab})} \ra \,
+ \, C_F C_A \, \la \Ph_{g_1 g_2 q_3}^{({\rm nab})} \ra  \;\;,
\eeq
and the abelian and non-abelian contributions are
\beeq
\label{qggabsf}
\la \Ph_{g_1 g_2 q_3}^{({\rm ab})} \ra \, 
&=&\Biggl\{\f{s_{123}^2}{2s_{13}s_{23}}
z_3\left[\f{1+z_3^2}{z_1z_2}-\ep\f{z_1^2+z_2^2}{z_1z_2}-\ep(1+\ep)\right]\nn\\
&+&\f{s_{123}}{s_{13}}\Biggl[\f{z_3(1-z_1)+(1-z_2)^3}{z_1z_2}+\ep^2(1+z_3)
-\ep (z_1^2+z_1z_2+z_2^2)\f{1-z_2}{z_1z_2}\Biggr]\nn\\
&+&(1-\ep)\left[\ep-(1-\ep)\f{s_{23}}{s_{13}}\right]
\Biggr\}+(1\lra 2) \;\;,
\eeeq
\beeq
\label{qggnabsf}
\la \Ph_{g_1 g_2 q_3}^{({\rm nab})} \ra \,
&=&\Biggl\{(1-\ep)\left(\f{t_{12,3}^2}{4s_{12}^2}+\f{1}{4}
-\f{\ep}{2}\right)+\f{s_{123}^2}{2s_{12}s_{13}}
\Biggl[\f{(1-z_3)^2(1-\ep)+2z_3}{z_2}\nn\\
&+&\f{z_2^2(1-\ep)+2(1-z_2)}{1-z_3}\Biggr]
-\f{s_{123}^2}{4s_{13}s_{23}}z_3\Biggl[\f{(1-z_3)^2(1-\ep)+2z_3}{z_1z_2}
+\ep(1-\ep)\Biggr]\nn\\
&+&\f{s_{123}}{2s_{12}}\Biggl[(1-\ep)
\f{z_1(2-2z_1+z_1^2) - z_2(6 -6 z_2+ z_2^2)}{z_2(1-z_3)}
+2\ep\f{z_3(z_1-2z_2)-z_2}{z_2(1-z_3)}\Biggr]\nn\\
&+&\f{s_{123}}{2s_{13}}\Biggl[(1-\ep)\f{(1-z_2)^3
+z_3^2-z_2}{z_2(1-z_3)}
-\ep\left(\f{2(1-z_2)(z_2-z_3)}{z_2(1-z_3)}-z_1 + z_2\right)\nn\\
&-&\f{z_3(1-z_1)+(1-z_2)^3}{z_1z_2}
+\ep(1-z_2)\left(\f{z_1^2+z_2^2}{z_1z_2}-\ep\right)\Biggr]\Biggr\}
+(1\lra 2) \;\;.
\eeeq

As discussed in Sect.~\ref{power},
in the case of collinear decays of a gluon (see Eqs.~(\ref{gqq}, \ref{ggg})),
spin correlations are highly non-trivial.

\begin{figure}[htb]
\begin{center}
\begin{tabular}{c}
\epsfxsize=13truecm
\epsffile{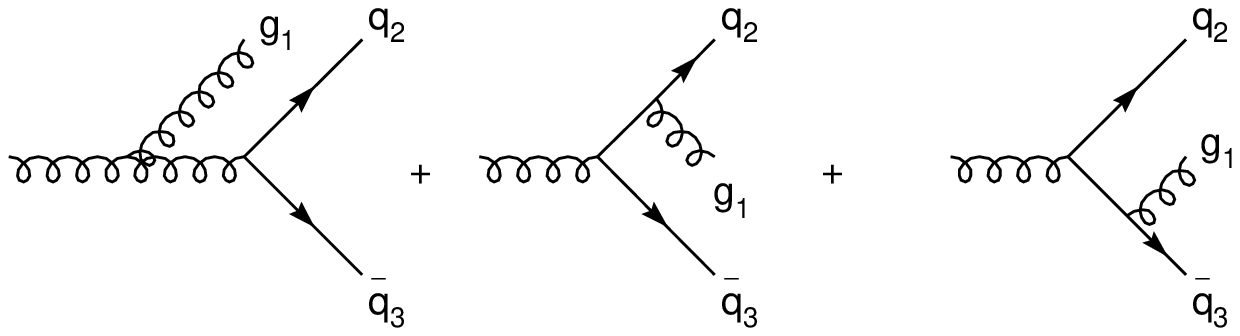}\\
\end{tabular}
\end{center}
\caption{\label{qqgfig}{\em The diagrams for the collinear decay
$g\to g_1q_2{\bar q}_3$. }}
\end{figure}

To compute the splitting function for the decay into a $q{\bar q}$ pair plus a 
gluon, we have to evaluate the square of the diagrams in Fig.~\ref{qqgfig}.
The colour-factor decomposition of the splitting function is
\beq
\label{gqqsf}
\Ph^{\mu\nu}_{g_1 q_2 {\bar q}_3}  \, =
C_F T_R \, \Ph_{g_1 q_2 {\bar q}_3}^{\mu\nu \,({\rm ab})} \,
+ \, C_A T_R\, \Ph_{g_1 q_2 {\bar q}_3}^{\mu\nu \,({\rm nab})}  \;\;,
\eeq
where the abelian and non-abelian terms are given by
\beeq
\label{gqqabsf}
\Ph^{\mu\nu \,({\rm ab})}_{g_1q_2{\bar q}_3} &=&
-g^{\mu\nu}\Biggl[ -2 
+ \f{2 s_{123} s_{23} + (1-\ep) (s_{123} - s_{23})^2}{s_{12}s_{13}}\Biggr]\nn\\
&+& \f{4s_{123}}{s_{12}s_{13}}\left({\ktil}_{3}^\mu
    {\ktil}_{2}^\nu+{\ktil}_{\hs 2}^\mu
    {\ktil}_{3}^\nu-(1-\ep){\ktil}_{\hs 1}^\mu 
    {\ktil}_{1}^\nu \right)
\;\;,
\eeeq
\beeq
\label{gqqnabsf}
\Ph^{\mu\nu \,({\rm nab})}_{g_1q_2{\bar q}_3} &=& \f{1}{4}
\,\Biggl\{ \f{s_{123}}{s_{23}^2}
\Biggl[ g^{\mu\nu} \f{t_{23,1}^2}{s_{123}}-16\f{z_2^2z_3^2}{z_1(1-z_1)}
\left(\f{{\ktil}_2}{z_2}-\f{{\ktil}_3}{z_3}\right)^\mu
\left(\f{{\ktil}_2}{z_2}-\f{{\ktil}_3}{z_3}\right)^\nu \,\Biggr]\nn\\
&+& \f{s_{123}}{s_{12}s_{13}} \Biggl[ 2 s_{123} g^{\mu\nu}
- 4 ( {\ktil}_2^\mu {\ktil}_3^\nu + {\ktil}_3^\mu {\ktil}_2^\nu
- (1-\ep) {\ktil}_1^\mu {\ktil}_1^\nu ) \Biggr] \nn\\
&-& g^{\mu\nu} \Biggl[ - ( 1 -2 \ep) + 2\f{s_{123}}{s_{12}} 
\f{1-z_3}{z_1(1-z_1)} + 2\f{s_{123}}{s_{23}} 
\f{1-z_1 + 2 z_1^2}{z_1(1-z_1)}\Biggr]\nn\\
&+& \f{s_{123}}{s_{12}s_{23}} \Biggl[ - 2 s_{123} g^{\mu\nu}
\f{z_2(1-2z_1)}{z_1(1-z_1)} - 16 {\ktil}_3^\mu {\ktil}_3^\nu 
\f{z_2^2}{z_1(1-z_1)} 
+ 8(1-\ep) {\ktil}_2^\mu {\ktil}_2^\nu \nn\\
&+& 4 ({\ktil}_2^\mu {\ktil}_3^\nu  + {\ktil}_3^\mu {\ktil}_2^\nu )
\left(\f{2 z_2 (z_3-z_1)}{z_1(1-z_1)}+ (1-\ep) \right)
\Biggr] \Biggr\} + \left( 2 \leftrightarrow 3 \right)  \;\;.
\eeeq

\begin{figure}[htb]
\begin{center}
\begin{tabular}{c}
\epsfxsize=12truecm
\epsffile{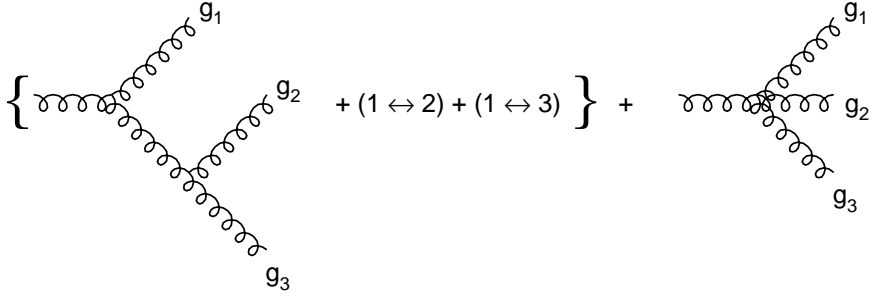}\\
\end{tabular}
\end{center}
\caption{\label{gggfig}{\em The diagrams for the collinear decay
$g\to g_1g_2g_3$. }}
\end{figure}

In the case of gluon decay into three collinear gluons we have to consider
the diagrams in Fig.~\ref{gggfig}. Note that the contribution of the
four-gluon vertex cannot be neglected. The result for the splitting function
is quite involved. Its expression can be simplified by exploiting the complete
symmetry under the six permutations of the gluon momenta. We obtain
\beeq
\label{gggsf}
\Ph^{\mu\nu}_{g_1g_2g_3} &=& C_A^2 
\,\Biggl\{\f{(1-\ep)}{4s_{12}^2}
\Biggl[-g^{\mu\nu} t_{12,3}^2+16s_{123}\f{z_1^2z_2^2}{z_3(1-z_3)}
\left(\f{{\ktil}_2}{z_2}-\f{{\ktil}_1}{z_1}\right)^\mu
\left(\f{{\ktil}_2}{z_2}-\f{{\ktil}_1}{z_1}\right)^\nu \;\Biggr]\nn\\
&-& \f{3}{4}(1-\ep)g^{\mu\nu}+\f{s_{123}}{s_{12}}g^{\mu\nu}\f{1}{z_3}
    \Biggl[\f{2(1-z_3)+4z_3^2}{1-z_3}-\f{1-2z_3(1-z_3)}{z_1(1-z_1)}\Biggr]\nn\\
&+& \f{s_{123}(1-\ep)}{s_{12}s_{13}}\Biggl[2z_1\left({\ktil}^\mu_2
    {\ktil}^\nu_2\hs\f{1-2z_3}{z_3(1-z_3)}+
    {\ktil}^\mu_3{\ktil}^\nu_3\hs
    \f{1-2z_2}{z_2(1-z_2)}\right)\nn\\
&+& \f{s_{123}}{2(1-\ep)} g^{\mu\nu}
    \left(\f{4z_2z_3+2z_1(1-z_1)-1}{(1-z_2)(1-z_3)}
    - \f{1-2z_1(1-z_1)}{z_2z_3}\right)\nn\\
&+& \left({\ktil}_2^\mu{\ktil}_3^\nu
   +{\ktil}_3^\mu{\ktil}_2^\nu\right)
    \left(\f{2z_2(1-z_2)}{z_3(1-z_3)}-3\right)\Biggr]\Biggr\}
    + (5\mbox{ permutations}) \;\;.
\eeeq

The splitting functions in Eqs.~(\ref{gqqabsf})--(\ref{gggsf})
can be averaged over the spin polarizations of the parent
gluon according to Eq.~(\ref{gsfav}):
\beq
\la \Ph_{a_1 a_2 a_3} \ra \, \equiv \frac{1}{2 (1 - \ep)} \,d_{\mu \nu}(p)
\; \Ph^{\mu\nu}_{a_1 a_2 a_3} \;\;.
\eeq
Performing the average, we obtain
\beeq
\label{gqqabsfav}
\la \Ph^{({\rm ab})}_{g_1q_2{\bar q}_3} \ra \,&=&
-2-(1-\ep)s_{23}\left(\f{1}{s_{12}}+\f{1}{s_{13}}\right)
+ 2\f{s_{123}^2}{s_{12}s_{13}}\left(1+z_1^2-\f{z_1+2z_2 z_3}{1-\ep}\right) 
\nn\\
&-&\f{s_{123}}{s_{12}}\left(1+2z_1+\ep-2\f{z_1+z_2}{1-\ep}\right)
- \f{s_{123}}{s_{13}}\left(1+2z_1+\ep-2\f{z_1+z_3}{1-\ep}\right)
\;,
\eeeq
\beeq
\label{gqqnabsfav}
\la \Ph^{({\rm nab})}_{g_1q_2{\bar q}_3} \ra
\,&=&\Biggl\{-\f{t^2_{23,1}}{4s_{23}^2}
+\f{s_{123}^2}{2s_{13}s_{23}} z_3
\Biggl[\f{(1-z_1)^3-z_1^3}{z_1(1-z_1)}
-\f{2z_3\left(1-z_3 -2z_1z_2\right)}{(1-\ep)z_1(1-z_1)}\Biggr]\nn\\
&+&\f{s_{123}}{2s_{13}}(1-z_2)\Biggl[1
+\f{1}{z_1(1-z_1)}-\f{2z_2(1-z_2)}{(1-\ep)z_1(1-z_1)}\Biggr]\nn\\
&+&\f{s_{123}}{2s_{23}}\Biggl[\f{1+z_1^3}{z_1(1-z_1)}
+\f{z_1(z_3-z_2)^2-2z_2z_3(1+z_1)}
{(1-\ep)z_1(1-z_1)}\Biggr] \nn\\
&-&\f{1}{4}+\f{\ep}{2}
-\f{s_{123}^2}{2s_{12}s_{13}}\Biggl(1+z_1^2-\f{z_1+2z_2z_3}{1-\ep}
\Biggr) \Biggr\}
+ (2\lra  3) \;\;,
\eeeq
\beeq
\label{gggsfav}
\la \Ph_{g_1g_2g_3} \ra \,&=& C_A^2\Biggl\{\f{(1-\ep)}{4s_{12}^2}
t_{12,3}^2+\f{3}{4}(1-\ep)+\f{s_{123}}{s_{12}}\Biggl[4\f{z_1z_2-1}{1-z_3}
+\f{z_1z_2-2}{z_3}+\f{3}{2} +\f{5}{2}z_3\nn\\
&+&\f{\left(1-z_3(1-z_3)\right)^2}{z_3z_1(1-z_1)}\Biggr]
+\f{s_{123}^2}{s_{12}s_{13}}\Biggl[\f{z_1z_2(1-z_2)(1-2z_3)}{z_3(1-z_3)}
+z_2z_3 -2 +\f{z_1(1+2z_1)}{2}\nn\\
&+&\f{1+2z_1(1+z_1)}{2(1-z_2)(1-z_3)}
+\f{1-2z_1(1-z_1)}{2z_2z_3}\Biggr]\Biggr\}
+ (5\mbox{ permutations}) \;\;.
\eeeq

The ${\cal O}(\as^2)$-collinear limit of tree-level QCD
amplitudes has been independently considered by Campbell and Glover 
[\ref{glover}]. They have computed only the spin-averaged splitting
functions. The comparison with their results has been discussed in detail in
Ref.~[\ref{lett}] and we do not repeat it here. 
Our results agree with those of Ref.~[\ref{glover}].
Since the methods used by the two groups are completely different
(cf. the discussion in Sect.~\ref{intro}),
this agreement can be regarded as an important cross-check of the calculations. 

The expressions of the spin-dependent splitting functions
${\hat P}^{s s'}_{a_1a_2a_3}$ derived in this section refer to the CDR scheme.
Other dimensional-regularization schemes can be used. We mention two of them,
which differ from CDR only by the number of spin-polarizations of quarks and
gluons.

The dimensional-reduction (DR) scheme [\ref{dimred}] works by considering two
spin-polarization states for quarks and two for gluons. The corresponding
spin-dependent splitting functions ${\hat P}^{s s'}_{a_1a_2a_3}$ are simply 
obtained from those in the CDR scheme by setting $\epsilon = 0$.

The `toy' dimensional-regularization (TDR) scheme introduced in 
Ref.~[\ref{schemedep}] considers $d-2=2(1-\epsilon)$ spin-polarization states 
for quarks as for gluons. Its practical implementation is very simple.
When computing traces of gamma matrices, we should use the relation
${\rm Tr} \;{\bom 1}= 4(1-\epsilon)$, where ${\bom 1}$ is the unity matrix in
the spinor space. The corresponding
spin-dependent splitting functions ${\hat P}^{s s'}_{a_1a_2a_3}$ are
obtained from those in the CDR scheme by the simple replacement 
$T_R \to T_R(1-\epsilon)$. 

The QCD results presented in this section can also be extended in a 
straightforward way to the abelian and supersymmetric cases.

In the case of QED, we have to perform the replacement $\as \to \alpha$ in the
factorization formula (\ref{ccfacm}), and the relevant splitting functions,
${\hat P}^{(QED)}_{a_1a_2a_3}$, for the triple collinear limit are obtained 
from the QCD splitting functions as
\beeq
{\hat P}^{(QED)}_{{\bar q}^{\prime}_1 q^{\prime}_2 q_3} &=& 
e_q^2 e_{q^\prime}^2 \Bigl( {\hat P}_{{\bar q}^{\prime}_1 q^{\prime}_2 q_3}
\Bigr)_{\rm {ab.}} \;\;, \nn \\
{\hat P}^{(QED)}_{{\bar q}_1 q_2 q_3} &=& e_q^4 \Bigl(
{\hat P}_{{\bar q}_1 q_2 q_3} \Bigr)_{\rm {ab.}} \;\;,\\
{\hat P}^{(QED)}_{\gamma_1 \gamma_2 q_3} &=& e_q^4 \Bigl(
{\hat P}_{g_1 g_2 q_3} \Bigr)_{\rm {ab.}} \;\;, \nn \\
{\hat P}^{(QED)}_{\gamma_1 q_2 {\bar q}_3} &=& e_q^4 \Bigl(
{\hat P}_{g_1 q_2 {\bar q}_3} \Bigr)_{\rm {ab.}} \;\;, \nn
\eeeq
where $e_q$ is the quark electric charge and the notation 
$\Bigl( \dots \Bigr)_{\rm {ab.}}$ means that we have to set $C_F=T_R=1$ and
$C_A=0$ in the QCD expression inside the round bracket.

The supersymmetric version of QCD, namely $N=1$ supersymmetric Yang--Mills 
theory, is obtained by replacing the quark with the gluino $\tilde g$,
a Majorana fermion in the adjoint representation of the gauge group.
To obtain the corresponding splitting functions, we have to change the colour
factors accordingly, and we have to identify $q={\bar q}=\tilde g$ after having
summed over the different permutations of the final-state fermions. We have
\beeq
{\hat P}_{{\tilde g}_1 {\tilde g}_2 {\tilde g}_3} &=& \Bigl(
{\hat P}_{{\bar q}_1 q_2 q_3} + 
{\hat P}_{q_1 {\bar q}_2 q_3} + {\hat P}_{q_1 q_2 {\bar q}_3}
\Bigr)_{\rm {SQCD}} \;\;, \nn \\
{\hat P}_{g_1 g_2 {\tilde g}_3} &=& \Bigl(
{\hat P}_{g_1 g_2 q_3} \Bigr)_{\rm {SQCD}} \;\;, \\
{\hat P}_{g_1 {\tilde g}_2 {\tilde g}_3} &=& \Bigl(
{\hat P}_{g_1 q_2 {\bar q}_3} + {\hat P}_{g_1 {\bar q}_2 q_3}
\Bigr)_{\rm {SQCD}} \;\;, \nn
\eeeq
where the notation $\Bigl( \dots \Bigr)_{\rm {SQCD}}$ means that we have to set
$C_F=2 T_R= C_A$ in the QCD expression inside the round bracket.

Gluino and gluon amplitudes are related by supersymmetry transformations.
In the collinear limit, these transformations relate 
the total splitting functions ${\hat P}^{s s'}_{{\tilde g} \to 3}$
and ${\hat P}^{\mu \nu}_{g \to 3}$ for gluino and gluon decays, which are
defined as
\beq
{\hat P}^{s s'}_{{\tilde g} \to 3} \equiv
{\hat P}^{s s'}_{{\tilde g}_1 {\tilde g}_2 {\tilde g}_3}
+ \left[ {\hat P}^{s s'}_{g_1 g_2 {\tilde g}_3} + (3\!\lra
\! 1)+(3\!\lra\! 2) \right] = \delta^{s s'} 
\;\la {\hat P}_{{\tilde g} \to 3} \ra \;\;,
\eeq
\beq
{\hat P}^{\mu \nu}_{g \to 3} \equiv
{\hat P}^{\mu \nu}_{g_1 g_2 g_3}
+ \left[ {\hat P}^{\mu \nu}_{g_1 {\tilde g}_2 {\tilde g}_3}
+ (1\!\lra \! 2)+(1\!\lra\! 3) \right] \;\;.
\eeq
In the four-dimensional supersymmetric theory, gluon and gluino have the same 
decay probability.
Provided supersymmetry is not broken by the dimensional-regularization 
procedure, we thus
have the following supersymmetric Ward identity:
\beq
\label{susywi}
{\hat P}^{\mu \nu}_{g \to 3} = - g^{\mu \nu} 
\;\la {\hat P}_{{\tilde g} \to 3} \ra \;\;.
\eeq
Note that the Ward identity holds for the spin-dependent splitting
functions. Since spin correlations are absent in the gluino splitting function,
they cancel in the right-hand side of Eq.~(\ref{susywi}), and 
${\hat P}^{ss'}_{{\tilde g} \to 3}$ and ${\hat P}^{\mu \nu}_{g \to 3}$ 
differ only by the overall spin-factors $\delta^{ss'}$ and $-g^{\mu\nu}$.
As is well known, the splitting functions
${\hat P}^{ss'}_{{\tilde g} \to 2}$ and ${\hat P}^{\mu \nu}_{g \to 2}$ 
are related by a similar Ward identity at ${\cal O}(\as)$.

The identity (\ref{susywi}) is violated in the CDR scheme, 
because gluinos and gluons have a different number of spin-polarization
states. The Ward identity is recovered in the $\ep\to 0$ limit or,
equivalently, in the DR scheme, which is known to explicitly preserve 
supersymmetry. Our results for the spin-dependent splitting functions fulfil 
Eq.~(\ref{susywi}), and this is an important check of our calculation.
As pointed out in Ref.~[\ref{schemedep}], the Ward identity at ${\cal O}(\as)$
is fulfilled also in the TDR scheme.
We have verified that this remains true at ${\cal O}(\as^2)$, as expected from
the fact that in the TDR scheme
the number of gluino states is the same as 
the number of gluon states.

\section{The soft behaviour}
\label{secsoft}

The tree-level matrix elements $\cm(p_1,p_2,\dots)$ are singular not only 
when parton momenta become collinear but also when one or more of them become 
soft. In QCD calculations of physical cross sections at NLO, the soft limit
is approached when the momentum of a single gluon vanishes. At NNLO we have to
consider three different types of soft configurations:
\begin{itemize}
\item the emission of a soft quark--antiquark pair,
\item the emission of two soft gluons,
\item the emission of a soft gluon and a pair of collinear partons.
\end{itemize}
The behaviour of the tree-level matrix elements in these
singular limits is considered in this section. We also discuss
the generalization of the corresponding factorization formulae
to higher perturbative orders.

\subsection{Colour correlations and eikonal current at ${\cal O}(\as)$}
\label{secsoftlo}

The emission of a soft gluon does not affect the kinematics (momenta and spins)
of the radiating partons. However, it does affect their colour because 
the gluon always carries away some colour charge, no matter how soft it is. 
Unlike the case of soft-photon emission in QED, soft-gluon emission thus
does not factorize exactly and leads to colour correlations.

To take into account the colour structure (as well as the spin and flavour
structures), it is useful to introduce a basis
$\{ \ket{c_1,...,c_n} \otimes \ket{s_1,...,s_n} \}$
in colour + helicity space in such a way that the tree-level matrix element
in Eq.~(\ref{meldef}) with $n$ final-state partons can be written as
\beq
\label{cmmdef}
\cm_{a_1,\dots,a_n}^{c_1,\dots,c_n; s_1,\dots,s_n}(p_1,\dots,p_n) \equiv
\Bigl( \bra{c_1,\dots,c_n} \otimes \bra{s_1,\dots,s_n} \Bigr) \;
\ket{\cm_{a_1,\dots,a_n}(p_1,\dots,p_n)} \;.
\eeq
Thus $\ket{\cm_{a_1,\dots,a_n}(p_1,\dots,p_n)}$
is a vector in colour + helicity space.

According to this notation, the matrix element squared (summed
over final-state colours and spins) $|\cm|^2$ can be written as
\beq
|\cm_{a_1,\dots,a_n}(p_1,\dots,p_n)|^2 = 
\la \, \cm_{a_1,\dots,a_n}(p_1,\dots,p_n) \,
 | \, \cm_{a_1,\dots,a_n}(p_1,\dots,p_n) \, \ra \;.
\eeq

To describe the colour correlations produced by soft-gluon emission,
it is convenient to associate a colour charge ${\bom T}_i$
with the emission of a gluon from each parton $i$. If the emitted gluon
has colour index $c$ ($c= 1, ...,$ $N_c^2-1$), the colour-charge operator is:
\beq
{\bom T}_i \equiv \bra{c} \;T_i^c 
\eeq
and its action onto the colour space is defined by
\beq
\bra{c_1,\dots, c_i, \dots, c_m, c} {\bom T}_i
\ket{b_1, \dots, b_i, \dots, b_m} = \delta_{c_1 b_1} ...
T_{c_i b_i}^c ...\delta_{c_m b_m} \;\;,
\eeq
where $T_{c b}^a \equiv i f_{cab}$ (colour-charge matrix
in the adjoint representation)  if the emitting particle $i$
is a gluon and $T_{\alpha \beta}^a \equiv t^a_{\alpha \beta}$
(colour-charge matrix in the fundamental representation with
$\alpha, \beta =1,\dots,N_c$)
if the emitting particle $i$ is a quark (in the case of an emitting
antiquark $T_{\alpha \beta}^a \equiv {\bar t}^a_{\alpha \beta}
= - t^a_{\beta \alpha }$).

The colour-charge algebra is\footnote{More details on the colour algebra 
and useful colour-matrix relations can be found in Appendix~A of 
Ref.~[\ref{CSdipole}].}:
\beq
T_i^c \, T_j^c \equiv
{\bom T}_i \cdot {\bom T}_j ={\bom T}_j \cdot {\bom T}_i \;\;\;\;{\rm if}
\;\;i \neq j; \;\;\;\;\;\;{\bom T}_i^2= C_i,
\eeq
where $C_i$ is the Casimir operator, that is,
$C_i=C_A=N_c$ if $i$ is a gluon and $C_i=C_F=(N_c^2-1)/2N_c$ if $i$ is a quark
or antiquark. 

Note that, by definition, each vector $\ket{\cm_{a_1,\dots,a_n}(p_1,\dots,p_n)}$
is a colour-singlet state. Therefore colour conservation is simply
\beq \label{cocon}
\sum_{i=1}^n {\bom T}_i \; \ket{\cm_{a_1,\dots,a_n}(p_1,\dots,p_n)} = 0 \;.
\eeq

Let us now consider the tree-level matrix element
${\cal M}_{g,a_1,\dots,a_n}(q,p_1,\dots,p_n)$ in the limit where the 
momentum $q$ of the gluon becomes soft. Denoting by
$c$ and $\mu$ the colour and spin indices of the soft gluon,
the matrix element fulfils the following factorization formula~[\ref{BCM}]
\beq
\label{eikfac}
\la c; \mu |\,{\cal M}_{g,a_1,\dots,a_n}(q,p_1,\dots,p_n) \rangle 
\simeq g_S \mu^\ep
J^{c;\mu}(q)
\; |\,{\cal M}_{a_1,\dots,a_n}(p_1,\dots,p_n)\rangle \;,
\eeq
where $|\,{\cal M}_{a_1,\dots,a_n}(p_1,\dots,p_n)\rangle$
is obtained from the original matrix by simply removing the soft gluon $q$.
The factor ${\bom J}^\mu(q)$ is the eikonal current
\beq
\label{eikcur}
{\bom J}^\mu(q)=\sum_{i=1}^{n} {\bom T}_i\,\f{p_i^\mu}{p_i\cdot q} \;,
\eeq
which depends on the momenta and colour charges of the
hard partons in the matrix element on the right-hand side of 
Eq.~(\ref{eikfac}). The symbol `$\simeq$' means that on the right-hand side
we have neglected contributions that are less singular than $1/q$ in the soft
limit $q \to 0$. Note that Eq.~(\ref{eikfac}) is valid in any number
$d=4-2\ep$ of space-time dimensions, and the sole dependence on $d$ is in the 
overall factor $\mu^\ep$.

The factorization formula (\ref{eikfac}) can be derived in a simple way
by working in a physical gauge and using the following
{\em soft-gluon insertion rules}. The coupling of the gluon to any 
{\em internal} (i.e. highly off-shell) parton in the amplitude 
${\cal M}_{g,a_1,\dots,a_n}(q,p_1,\dots,p_n)$ is not singular in the soft limit;
it can thus be neglected. The soft-gluon coupling to any {\em external}
or, in general, {\em nearly on-shell} parton with colour charge $\bom T$
and momentum $p$ can be factorized by extracting the contribution 
$g_S \mu^\ep 2 p^\mu {\bom T}$ for the vertex and the contribution 
$1/(p+q)^2 \simeq 1/(p^2 + 2p\cdot q)$ for the propagator.

An important property of the eikonal current is current conservation. 
Multiplying Eq.~(\ref{eikfac}) by $q^\mu$, we obtain
\beq
\label{eikcons}
q_\mu {\bom J}^\mu(q) = \sum_{i=1}^{n} {\bom T}_i \;,
\eeq
and thus
\beq
q_\mu {\bom J}^\mu(q) |\,{\cal M}_{a_1,\dots,a_n}(p_1,\dots,p_n)\rangle
= \sum_{i=1}^{n} {\bom T}_i \;|\,{\cal M}_{a_1,\dots,a_n}(p_1,\dots,p_n)\rangle
= 0 \;\;,
\eeq
where the last equality follows from colour conservation 
as in Eq.~(\ref{cocon}).

Although Eq.~(\ref{eikfac}) is most easily derived in a physical gauge,
the conservation of the eikonal current implies the gauge invariance
of the squared amplitude summed over the soft-gluon polarizations.
Squaring the eikonal current and introducing the gluon
polarization tensor $d_{\mu \nu}(q) = ( - g_{\mu \nu} + {\rm gauge \; terms})$ 
in Eq.~(\ref{gluav}), we have
\beq
\label{eikonal2}
\left[ {\bom J}^{\mu}(q) \right]^\dagger \;d_{\mu \nu}(q) \;{\bom J}^{\nu}(q) =
- \sum_{i,j=1}^n \;{\bom T}_i \cdot {\bom T}_j 
\;\frac{p_i \cdot p_j}{(p_i \cdot q) (p_j \cdot q)} + \dots \;\;,
\eeq
where we have used the fact that the gauge terms in $d_{\mu \nu}(q)$ are due to 
longitudinal polarizations proportional either to $q^\mu$ or to $q^\nu$. Thus,
the dots on the right-hand side stand for gauge-dependent contributions that 
are proportional to the total colour charge $\sum_{i=1}^{n} {\bom T}_i$ and,
hence, that cancel when they are inserted in 
$|\,{\cal M}_{a_1,\dots,a_n}(p_1,\dots,p_n)\rangle$.

Using Eq.~(\ref{eikonal2}), the soft-gluon factorization formula at 
${\cal O}(\as)$ for the squared amplitude~is
\beeq
\label{ccfact}
| \cm_{g,a_1,\dots,a_n}(q,p_1,\dots,p_n) |^2 \simeq
- 4 \pi \as \mu^{2\ep}  \sum_{i,j=1}^n\, {\cal S}_{ij}(q) 
\;| \cm^{(i,j)}_{a_1,\dots,a_n}(p_1,\dots,p_n) |^2 \;\;,
\eeeq 
where the scalar eikonal function ${\cal S}_{ij}(q)$ for the emission of
a single gluon can be written in terms of two-particle sub-energies 
$s_{ij}=(p_i + p_j)^2$ as follows
\beq
\label{eikfun}
{\cal S}_{ij}(q) = \f{p_i \cdot p_j}{(p_i \cdot q)\, (p_j\cdot q)}
= \f{ 2 s_{ij}}{s_{iq} \,s_{jq}} \;\;.
\eeq
The colour correlations produced by soft-gluon emission are taken into
account by the square of the colour-correlated tree-amplitude 
$| \cm^{(i,j)} |^2$ on the right-hand side. This is defined by
\beeq
\label{colam}
| \cm^{(i,j)}_{a_1,\dots,a_n}(p_1,\dots,p_n) |^2 \!\!\!&\equiv&\!\!\!
\la \,{\cal M}_{a_1,\dots,a_n}(p_1,\dots,p_n) \,|
\,{\bom T}_i \cdot {\bom T}_j 
\,|\,{\cal M}_{a_1,\dots,a_n}(p_1,\dots,p_n)\rangle
\\
\!&=&\!\!\! \left[ 
{\cal M}_{a_1,\dots,a_n}^{c_1.. b_i ... b_j ... c_n}(p_1,\dots,p_n) \right]^*
\; T_{b_id_i}^c \, T_{b_jd_j}^c
\; {\cal M}_{a_1,\dots,a_n}^{c_1.. d_i ... d_j ... c_n}(p_1,\dots,p_n) \;,
\nonumber
\eeeq
where the sum over the spin indices is understood.

The soft-gluon factorization formula is often presented [\ref{glover}]
in an equivalent way by decomposing the matrix element in terms of 
colour subamplitudes [\ref{mangano}]. In this formalism, the eikonal function 
${\cal S}_{ij}(q)$ in Eq.~(\ref{eikfun}) controls the factorization properties
of the square of the colour-connected subamplitudes.

\subsection{Emission of a soft $q{\bar q}$-pair}
\label{secsoftqq}

We now consider the tree-level matrix element 
$\cm_{q,{\bar q},a_1,\dots,a_n}(q_1,q_2,p_1,\dots,p_n)$ when the momenta 
$q_1$ and $q_2$ of the quark $q$ and the antiquark ${\bar q}$ become soft 
($q_1,q_2\to 0$ at
fixed $q_1/q_2$). In this limit the matrix element squared has the
dominant behaviour:
\beq
| \cm_{q,{\bar q},a_1,\dots,a_n}(q_1,q_2,p_1,\dots,p_n) |^2\sim
\f{1}{(q_1 \cdot q_2)\, [p_i \cdot (q_1+q_2)]\, [p_j \cdot (q_1+q_2)]} \;\;.
\eeq
When integrated over the phase space of
the quark-antiquark pair, this behaviour gives rise to a single-logarithmic
soft singularity, in addition to possible single- and double-logarithmic
collinear singularities.

\begin{figure}[htb]
\begin{center}
\begin{tabular}{c}
\epsfxsize=10truecm
\epsffile{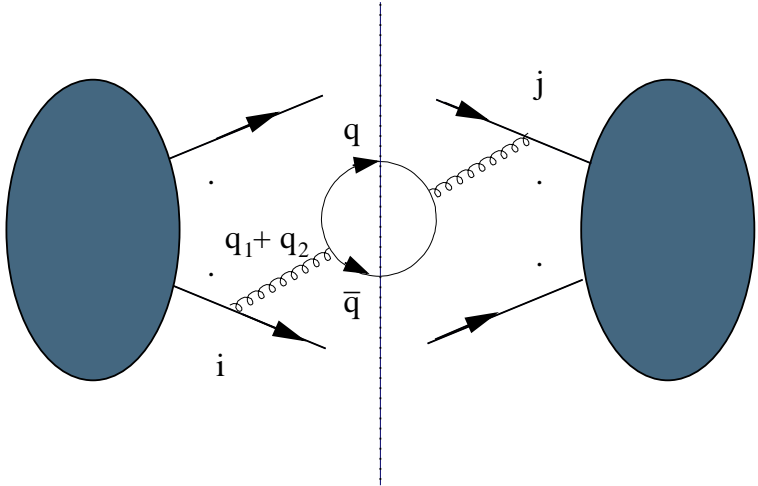}\\
\end{tabular}
\end{center}
\caption{\label{figsoftqq}{\em Soft-gluon insertion diagram for the emission
of a soft $q{\bar q}$ pair.
}}
\end{figure}

The soft singularity arises when the $q{\bar q}$-pair is produced by the decay
of a gluon that carries the soft momentum $q_1+q_2$ (Fig.~\ref{figsoftqq}). 
Thus, using the soft-gluon insertion rules described in Sect.~\ref{secsoftlo},
we can straightforwardly derive the following factorization formula:
\beeq
\label{insqq}
&&| \cm_{q,{\bar q},a_1,\dots,a_n}(q_1,q_2,p_1,\dots,p_n) |^2 \simeq
(4 \pi \mu^{2\ep} \as)^2\nn\\ 
&&\;\;\;\;\;\;\;\;\;\;\;\;\;\;\;\;\;\;\;\;\;\;\;\;\;\;\; 
\cdot \;
\langle \,{\cal M}_{a_1,\dots,a_n}(p_1,\dots,p_n) \,|
\; {\bom I}_{(q{\bar q})}(q_1,q_2) \;
| \, {\cal M}_{a_1,\dots,a_n}(p_1,\dots,p_n) \,\rangle 
\;, 
\eeeq
where
\beeq
\label{ioper1}
{\bom I}_{(q{\bar q})}(q_1,q_2)&=&
\left[ {\bom J}^\lambda(q_1 + q_2) \right]^\dagger d_{\lambda \mu}(q_1 + q_2) 
\; \Pi^{\mu \nu}(q_1,q_2) \; d_{\nu \rho}(q_1 + q_2) 
\;{\bom J}^\rho(q_1 + q_2) \\
\label{ioper2}
&=& \left[ {\bom J}_\mu(q_1 + q_2) \right]^\dagger 
\; \Pi^{\mu \nu}(q_1,q_2) \;{\bom J}_\nu(q_1 + q_2) +~{\dots} \;\;.
\eeeq
The insertion operator ${\bom I}_{(q{\bar q})}(q_1,q_2)$ depends on the
colour charges of the fast partons $a_1,\dots,a_n$ and it is given in terms
of the soft-gluon current ${\bom J}^\mu(q_1 + q_2)$ in Eq.~(\ref{eikcur}) 
and of $\Pi^{\mu\nu}(q_1,q_2)$, which is the $q{\bar q}$-contribution 
to the discontinuity of the gluon propagator:
\beq
\Pi^{\mu\nu}(q_1,q_2) = \frac{T_R}{(q_1 \cdot q_2)^2} \;
\Bigl\{ \;- g^{\mu\nu} q_1 \cdot q_2 + q_1^\mu q_2^\nu + q_2^\mu q_1^\nu 
\; \Bigr\} \;\;.
\eeq

The dots on the right-hand side of Eq.~(\ref{ioper2}) denote 
the gauge-dependent contribution to the insertion operator
${\bom I}_{(q{\bar q})}(q_1,q_2)$.
This term is due to the longitudinal
polarizations (proportional to $(q_1 + q_2)^\alpha$ or $(q_1 + q_2)^\beta$)
of the polarization tensors $d_{\alpha \beta}(q_1 + q_2)$ in 
Eq.~(\ref{ioper1}). Since $\Pi^{\mu\nu}(q_1,q_2)$ is transverse 
$(q_{1 \mu} \Pi^{\mu\nu}(q_1,q_2) = 0)$ and the soft current 
${\bom J}^\mu(q_1 + q_2)$ is conserved (see Eq.~(\ref{eikcons})),
the contribution of the longitudinal 
polarizations is either vanishing or proportional to the total colour charge
of the fast partons. Because of colour conservation (see Eq.~(\ref{cocon})),
we thus conclude that the gauge-dependent part of 
${\bom I}_{(q{\bar q})}(q_1,q_2)$ does not contribute to the
factorization formula  (\ref{insqq}).

Inserting Eq.~(\ref{ioper2}) into Eq.~(\ref{insqq}) and performing the 
Lorentz algebra, we obtain the final factorization formula
\beq
\label{qqsoftfac}
| \cm_{q,{\bar q},a_1,\dots,a_n}(q_1,q_2,p_1,\dots,p_n) |^2 \simeq
 (4 \pi \mu^{2\ep} \as)^2 \,T_R 
\sum_{i,j=1}^{n} {\cal I}_{ij}(q_1,q_2) \,
| \cm_{a_1,\dots,a_n}^{(i,j)}(p_1,\dots,p_n) |^2 ,
\eeq
where
$| \cm_{a_1,\dots,a_n}^{(i,j)}|^2$ is the colour-correlated tree-amplitude
of Eq.~(\ref{colam}) and the soft function ${\cal I}_{ij}(q_1,q_2)$ is given by
\beeq
\label{Iij1}
{\cal I}_{ij}(q_1,q_2)&=& \f{(p_i \cdot q_1)\, (p_j \cdot q_2)
+ (p_j \cdot q_1)\, (p_i \cdot q_2) - (p_i \cdot p_j) 
\,(q_1 \cdot q_2)}{(q_1 \cdot q_2)^2 
\,[p_i\cdot (q_1+q_2)]\, [p_j \cdot (q_1+q_2)]} \\
\label{Iij2}
&=& - \;\f{2 (p_i \cdot p_j) \,(q_1 \cdot q_2)
+ [p_i \cdot (q_1 - q_2)]\, [p_j \cdot (q_1 - q_2)]}{2 (q_1 \cdot q_2)^2 
\,[p_i\cdot (q_1+q_2)]\, [p_j \cdot (q_1+q_2)]} + \dots \,.
\eeeq
Note that both the expressions  (\ref{Iij1}) and (\ref{Iij2}) can equivalently
be used to compute Eq.~(\ref{qqsoftfac}). The difference between the two 
expressions, denoted by the dots on the right-hand side of Eq.~(\ref{Iij2}),
gives a vanishing contribution to the factorization formula (\ref{qqsoftfac})
because of colour conservation (see Eq.~(\ref{cocon})).

\subsection{Soft current for double gluon emission}
\label{secsoftgg}

The limit of QCD tree-amplitudes when the momenta
of two gluons simultaneously become soft was independently
studied by Berends and Giele [\ref{bgdsoft}] and by one of the authors
[\ref{sdsoft}]. The singular behaviour of the matrix elements
can be described in terms of factorization formulae given in terms
of process-independent two-gluon currents acting either on 
colour-ordered subamplitudes [\ref{bgdsoft}] or on the colour space of the
hard partons [\ref{sdsoft}]. 

The formalism of the colour subamplitudes was used in Ref.~[\ref{glover}] 
to derive explicit soft-gluon factors for the square of 
colour-connected and colour-unconnected subamplitudes. In this section
we recall the formalism and the results of Ref.~[\ref{sdsoft}] and we present
the corresponding factorization formula for the square of the matrix elements. 

We consider the tree-level matrix element 
${\cal M}_{g,g,a_1,\dots,a_n}(q_1,q_2,,p_1,\dots,p_n)$ when
the momenta $q_1$ and $q_2$ of the two gluons become soft. 
The limit is precisely defined by rescaling the gluon momenta by an
overall factor $\lambda$:
\beq
q_1\to \lambda q_1 \,, \;\;\;\;\;\; q_2\to \lambda q_2 \,,
\eeq
and then performing the limit $\lambda\to 0$.
The matrix element thus behaves as
\beq
\cm_{g,g,\dots}\to {\cal O}(1/\lambda^2)+\dots
\eeq
where the dots stand for less singular contributions as $\lambda\to 0$.
We are interested in explicitly evaluating the dominant singular term
${\cal O}(1/\lambda^2)$.

Note that the double soft limit is more general (accurate) than the soft limit
in the strong-ordering approximation, that is, when $p_i \gg q_1 \gg q_2$.
The strongly-ordered limit describes only the double-logarithmic
soft singularity of the matrix elements. The double soft limit  
reproduces consistently the double-logarithmic behaviour and correctly 
evaluates also the single-logarithmic soft singularity.

We denote by $a_1, a_2$ and $\mu_1, \mu_2$ the colour and Lorentz indices
of the two gluons, respectively. In the double soft limit the matrix element
fulfils the following factorization formula~[\ref{sdsoft}]
\beq
\label{softff2}
\bra{a_1,a_2;\mu_1, \mu_2} \,
\cm_{g,g,a_1,\dots,a_n}(q_1,q_2,p_1,\dots,p_n)\rangle 
\simeq g_S^2 \mu^{2\ep} \, J^{a_1a_2}_{\mu_1\mu_2}(q_1,q_2)\,
|\,\cm_{a_1,\dots,a_n}(p_1,\dots,p_n) \rangle \;,
\eeq
where the two-gluon soft current $J^{a_1a_2}_{\mu_1\mu_2}(q_1,q_2)$
is the generalization of the eikonal current in Eq.~(\ref{eikcur}).

\begin{figure}[htb]
\begin{center}
\begin{tabular}{c}
\epsfxsize=14truecm
\epsffile{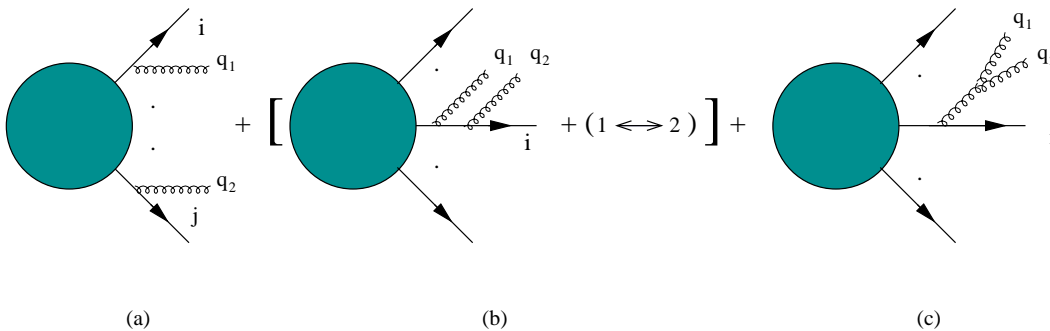}\\
\end{tabular}
\end{center}
\caption{\label{figdsoft}{\em Soft-gluon insertion diagrams used to evaluate
the two-gluon current $J_{a_1a_2}^{\mu_1\mu_2}(q_1,q_2)$.}}
\end{figure}

The explicit expression of the soft current is~[\ref{sdsoft}]
\beeq
\label{dsoftcur}
J_{a_1a_2}^{\mu_1\mu_2}(q_1,q_2)&=&\sum_{i \neq j}
T^{a_1}_i\f{p_i^{\mu_1}}{p_i \cdot q_1}\; 
T_j^{a_2}\f{p_j^{\mu_2}}{p_j\cdot q_2}+\nn\\
&+&\sum_i\Bigg[\left(\delta^{a_1a}\, T_i^{a_2}\, \f{p_i^{\mu_2}}{p_i\cdot q_2}
-if^{a_2a_1a}\f{q_1^{\mu_2}}{q_1\cdot q_2}\right)\, T_i^a\, 
\f{p_i^{\mu_1}}{p_i\cdot (q_1+q_2)}+\nn\\
&+&\left(\delta^{a_2a}\, T_i^{a_1}\, \f{p_i^{\mu_1}}{p_i\cdot q_1}
-if^{a_1a_2a}\f{q_2^{\mu_1}}{q_1\cdot q_2}\right)\,
T_i^a\, \f{p_i^{\mu_2}}{p_i\cdot (q_1+q_2)}+\nn\\
&+&\f{1}{2}if^{aa_1a_2}T_i^a\f{g^{\mu_1\mu_2}}{q_1\cdot q_2}
\;\f{p_i\cdot (q_2-q_1)}{p_i\cdot (q_2+q_1)}\Bigg].
\eeeq
It can be derived by working 
in a physical gauge and using the soft-gluon insertion technique described in 
Sect.~\ref{secsoftlo}. We have to consider the diagrams in Fig.~\ref{figdsoft}.
The contribution on the first line of Eq.~(\ref{dsoftcur})
comes from the eikonal emission of the two soft gluons from two
different external partons (diagrams $(\rm a)$ in Fig. \ref{figdsoft}). 
The first term on  
the second and third lines come from the eikonal emission of the two gluons 
from the same external parton (diagrams $(\rm b)$ in Fig.~\ref{figdsoft}). 
The remaining contributions in Eq.~(\ref{dsoftcur}) are proportional
to $f_{aa_1a_2}$ and originate from the 
non-abelian diagrams of Fig.~\ref{figdsoft}~(c). Note that the three-gluon
vertex has to be treated exactly, without introducing any soft approximation.

The two-gluon current in Eq.~(\ref{dsoftcur}) can be recast in the following 
equivalent form
\beeq
\label{dsoftcurac}
&&\!\!\!\!\!\!\!\! J_{a_1a_2}^{\mu_1\mu_2}(q_1,q_2)= \frac{1}{2} \left\{ 
J_{a_1}^{\mu_1}(q_1) \;, J_{a_2}^{\mu_2}(q_2) \right\} \\
&+& i f_{a_1a_2a} \sum_{i=1}^{n} T_i^a \left\{
\frac{p_i^{\mu_1} q_1^{\mu_2} - p_i^{\mu_2} 
q_2^{\mu_1}}{(q_1\cdot q_2) \,[p_i\cdot (q_1+q_2)]}
- \frac{p_i\cdot (q_1-q_2)}{2 [p_i\cdot (q_1+q_2)]}
\left[ \frac{p_i^{\mu_1} p_i^{\mu_2}}{(p_i \cdot q_1) (p_i \cdot q_2)} 
+ \f{g^{\mu_1\mu_2}}{q_1\cdot q_2} \right] \right\} \;, \nn
\eeeq
where the first term on the right-hand side is the colour anticommutator
of the single-gluon eikonal currents of Eq.~(\ref{eikcur}). This is the 
only contribution that survives in the abelian case, where it reduces itself
to the product of two independent single-gluon currents. The second term
on the right-hand side is typical of the non-abelian theory.

Note that, as in the single-gluon case, the expressions (\ref{dsoftcur})
and (\ref{dsoftcurac}) for the two-gluon current do not explicitly depend
on the number $d=4-2\epsilon$ of space-time dimensions. However, because
of the contribution proportional to $g^{\mu_1\mu_2}$ in 
$J_{a_1a_2}^{\mu_1\mu_2}$, an explicit dependence on the number
$d-2=2(1-\epsilon)$ of gluon polarizations appears (see Eqs.~(\ref{eik22})
and (\ref{dsoftfun}))
by squaring the factorization formula~(\ref{softff2}).

The current $J^{a_1a_2}_{\mu_1\mu_2}(q_1,q_2)$ fulfils the following
properties.
\begin{itemize}
\item It is symmetric under the exchange of the two soft gluons,
\begin{equation}
J^{a_1a_2}_{\mu_1\mu_2}(q_1,q_2)=J^{a_2a_1}_{\mu_2\mu_1}(q_2,q_1) \;.
\end{equation}
\item Its divergence is proportional to the total colour charge 
of the hard partons:
\begin{equation}
q_1^{\mu_1} J^{a_1a_2}_{\mu_1\mu_2}(q_1,q_2) =
\left( J^{a_2}_{\mu_2}(q_2) \;\delta_{a_1 a} + \frac{i}{2}
f_{a_1a_2a} \frac{q_1^{\mu_2}}{q_1\cdot q_2} \right) 
\sum_{i=1}^{n} T_i^a \;\;,
\end{equation}
\begin{equation}
q_2^{\mu_2} J^{a_1a_2}_{\mu_1\mu_2}(q_1,q_2) =
\left( J^{a_1}_{\mu_1}(q_1) \;\delta_{a_2 a} + \frac{i}{2}
f_{a_2a_1a} \frac{q_2^{\mu_1}}{q_1\cdot q_2} \right) 
\sum_{i=1}^{n} T_i^a \;\;.
\end{equation}
This property is analogous to Eq.~(\ref{eikcons}) for the single-gluon 
emission and implies that the two-gluon current is conserved 
when it acts on a colour singlet state:
\begin{equation}
q_1^{\mu_1} J^{a_1a_2}_{\mu_1\mu_2}(q_1,q_2)|\,\cm_{a_1,\dots}(p_1,\dots)
\rangle= q_2^{\mu_2} J^{a_1a_2}_{\mu_1\mu_2}(q_1,q_2)|
\,\cm_{a_1,\dots}(p_1,\dots)\rangle=0.
\end{equation}
Thus, the factorization formula (\ref{softff2}) is gauge-invariant.
\item In the strong-ordered limit $q_2 \ll q_1$, the third and fourth
  lines in Eq.~(\ref{dsoftcur}) give subleading contributions
and the current becomes
\begin{equation}
J_{a_1a_2}^{\mu_1\mu_2}(q_1,q_2) \rightarrow
\left( J_{a_2}^{\mu_2}(q_2) \;\delta_{a_1 a} + 
i f_{a_1a_2a} \frac{q_1^{\mu_2}}{q_1\cdot q_2} \right) 
J_{a}^{\mu_1}(q_1) \;\;.
\end{equation}
Thus, the current correctly factorizes into the product of the two eikonal
currents corresponding to the iterative application of the leading-order
factorization formula~(\ref{eikfac}).
\end{itemize}

The double soft limit of 
$| {\cal M}_{g,g,a_1,\dots,a_n}(q_1,q_2,,p_1,\dots,p_n) |^2$
is obtained by squaring Eq.~(\ref{softff2}) and by summing over the soft-gluon
polarizations. The square of the two-gluon current involves a quite cumbersome
colour algebra. Nonetheless, we find that the final result can be recast in 
a relatively simple form: 
\beeq
\!\!\! \left[J^{a_1a_2}_{\mu\rho}(q_1,q_2)\right]^\dagger
\,d^{\mu\nu}(q_1)\,d^{\rho\sigma}(q_2)\, J^{a_1a_2}_{\nu\sigma}(q_1,q_2)
&=& \f{1}{2}\left\{{\bom J}^2(q_1) \;, {\bom J}^2(q_2) \right\} \nn \\
\label{eik22}
&-& \,C_A\, \sum_{i,j=1}^{n} {\bom T}_i\cdot {\bom T}_j
\;{\cal S}_{ij}(q_1,q_2) + \dots \;,
\eeeq
where, as in Eq.~(\ref{eikonal2}), the dots stand for gauge-dependent terms.
These are proportional to the total colour charge of the hard partons and,
thus, give a vanishing contribution when
inserted on $|\,\cm_{a_1,\dots,a_n}(p_1,\dots,p_n)\rangle$.

The first term on the right-hand side of Eq.~(\ref{eik22}) is the only one that
survives in the abelian case. It is given in terms of the colour anticommutator
of the squares of the single-gluon currents in Eq.~(\ref{eikonal2}).
The second term is proportional to $C_A$ and, hence, is purely non-abelian.
It is given in terms of the two-gluon soft function ${\cal S}_{ij}(q_1,q_2)$:
\beeq
{\cal S}_{ij}(q_1,q_2) &=& \f{(1-\ep)}{(q_1 \cdot q_2 )^2} \;
\f{p_i \cdot q_1 \,p_j \cdot \,q_2 + p_i \cdot q_2 \,p_j \cdot \,q_1}
{p_i\cdot (q_1+q_2) \; p_j\cdot (q_1+q_2)} \nn \\
\label{dsoftfun}
&-& \f{(p_i \cdot p_j)^2}{2 p_i\cdot q_1 \; p_j\cdot q_2 \;
p_i\cdot q_2 \; p_j\cdot q_1}
\left[ 2 - \f{p_i \cdot q_1 \,p_j \cdot \,q_2 + p_i \cdot q_2 \,p_j \cdot \,q_1}
{p_i\cdot (q_1+q_2) \; p_j\cdot (q_1+q_2)} \right] \\
&+& \f{p_i\cdot p_j}{2 q_1 \cdot q_2}
\left[ \f{2}{p_i\cdot q_1 \,p_j \cdot \,q_2} + 
       \f{2}{p_j\cdot q_1 \,p_i \cdot \,q_2} \right. \nn \\
&-& \left.
       \f{1}{p_i\cdot (q_1+q_2) \; p_j\cdot (q_1+q_2)}
   \left( 4 + 
  \f{(p_i \cdot q_1 \,p_j \cdot \,q_2 + p_i \cdot q_2 
  \,p_j \cdot \,q_1)^2}{p_i\cdot q_1 \; p_j\cdot q_2 \;
  p_i\cdot q_2 \; p_j\cdot q_1}
\right) \right] \;\;. \nn 
\eeeq

Expression (\ref{dsoftfun}) can also be written as 
\beeq
{\cal S}_{ij}(q_1,q_2) &=& {\cal S}_{ij}^{{\rm (s.o.)}}(q_1,q_2)
+ \f{p_i \cdot q_1 \,p_j \cdot \,q_2 + p_i \cdot q_2 \,p_j \cdot \,q_1}
{p_i\cdot (q_1+q_2) \; p_j\cdot (q_1+q_2)}
\left[ \f{(1-\ep)}{(q_1 \cdot q_2 )^2} - \f{1}{2} 
\;{\cal S}_{ij}^{{\rm (s.o.)}}(q_1,q_2) \right] \nn \\
&-& 
\f{2 p_i\cdot p_j}{q_1 \cdot q_2 \;p_i\cdot (q_1+q_2) \; p_j\cdot (q_1+q_2) }
\;\;,
\eeeq
where ${\cal S}_{ij}^{{\rm (s.o.)}}$ is the approximation of the
soft function ${\cal S}_{ij}(q_1,q_2)$ in the strong-ordering 
limit (either $q_1 \ll q_2$ or $q_2 \ll q_1$):
\beeq
{\cal S}_{ij}^{{\rm (s.o.)}}(q_1,q_2) =
\f{p_i\cdot p_j}{q_1 \cdot q_2}
\left( \f{1}{p_i\cdot q_1 \,p_j \cdot \,q_2} + 
       \f{1}{p_j\cdot q_1 \,p_i \cdot \,q_2} \right)
-  \f{(p_i \cdot p_j)^2}{p_i\cdot q_1 \; p_j\cdot q_2 \;
p_i\cdot q_2 \; p_j\cdot q_1} \;\;.    
\eeeq

Using Eq.~(\ref{eik22}), we can write the 
soft-gluon factorization formula for the square of the matrix element 
as follows:
\beeq
\label{dsoftm2}
&&\!\!\!\!\!\!\!\! 
| {\cal M}_{g,g,a_1,\dots,a_n}(q_1,q_2,p_1,\dots,p_n) |^2 \simeq
\left( 4 \pi \as \mu^{2\epsilon} \right)^2  \\
&&\!\!\!\!\! \cdot \left[ \f{1}{2} \sum_{i,j,k,l=1}^{n} {\cal S}_{ij}(q_1) \;
{\cal S}_{kl}(q_2) \;
| {\cal M}_{a_1,\dots,a_n}^{(i,j)(k,l)}(p_1,\dots,p_n) |^2
- \,C_A\, \sum_{i,j=1}^{n} {\cal S}_{ij}(q_1,q_2)
| {\cal M}_{a_1,\dots,a_n}^{(i,j)} |^2
\right] \nn \;\;,
\eeeq
where ${\cal S}_{ij}(q)$ is the soft function in Eq.~(\ref{eikfun}) and
$| {\cal M}_{a_1,\dots,a_n}^{(i,j)} |^2$ is the colour-correlated amplitude
in Eq.~(\ref{colam}). We can see that the double soft limit 
involves colour correlations that are more cumbersome than those
appearing in the case of single-gluon emission. Indeed, the amplitude 
$| {\cal M}_{a_1,\dots,a_n}^{(i,j)(k,l)} |^2$ on the right-hand side
of Eq.~(\ref{dsoftm2}) is defined by
\beq
| {\cal M}_{a_1,\dots,a_n}^{(i,j)(k,l)}(p_1,\dots,p_n) |^2
\equiv
\langle \cm_{a_1,\dots,a_n}(p_1,\dots,p_n) | 
\left\{{\bom T}_i \cdot {\bom T}_j \;, {\bom T}_k \cdot {\bom T}_l \right\}
| \cm_{a_1,\dots,a_n}(p_1,\dots,p_n) \rangle \;,
\eeq
and leads to irreducible correlations among four different hard partons.

The results discussed in this subsection can be  
presented in a different manner by using the colour subamplitude formalism.
Considering the projection of Eq.~(\ref{softff2}) onto colour-ordered 
subamplitudes, it is straightforward to check that the colour current
$J_{a_1a_2}^{\mu_1\mu_2}(q_1,q_2)$ leads to the colourless current derived by
Berends and Giele (see Eqs.~(3.11) and (3.18) in Ref.~[\ref{bgdsoft}]).
The square of this colourless current is denoted by $S_{iq_1q_2j}$ in 
Sect.~5.3 of Ref.~[\ref{glover}] and is related to the soft function
${\cal S}_{ij}(q_1,q_2)$ in Eq.~(\ref{dsoftfun}). More
precisely, using the following relation
\beq
\label{dsoftvscg}
\sum_{i,j=1}^{n} {\bom T}_i\cdot {\bom T}_j \left[
{\cal S}_{ij}(q_1,q_2) + {\cal S}_{ij}(q_1) \;{\cal S}_{ij}(q_2) \right]
= \f{1}{2} \sum_{i,j=1}^{n} {\bom T}_i\cdot {\bom T}_j \;S_{iq_1q_2j} 
+ \dots \;\;,
\eeq
the second term on the right-hand side of Eqs.~(\ref{eik22})
and (\ref{dsoftm2}) can equivalently be written in terms
of $S_{iq_1q_2j}$. The contribution denoted by the dots on the right-hand side
of Eq.~(\ref{dsoftvscg}) is proportional to the total colour charge of the hard
partons and, thus, it vanishes when inserted in the factorization formula 
(\ref{dsoftm2}).

\subsection{Soft--collinear limit at ${\cal O}(\as^2)$ and at higher orders}
\label{softcoll}

We now consider the tree-level matrix element
${\cal M}_{g,a_1,\dots,a_n}(q,p_1,\dots,p_n)$
in the limit where the momentum $q$ of the gluon becomes soft $( q \to 0)$
and, at the same time, two partons, say $p_1$ and $p_2$, become
collinear. The collinear region is parametrized as in Eq.~(\ref{clim})
and we are interested in the limit $k_\perp \to 0$.

Studying this soft--collinear limit we can
neglect $i)$ contributions that are uniformly of ${\cal O}(q)$ when $q \to 0$, 
and $ii)$ contributions that are uniformly of ${\cal O}(k_\perp)$ when 
$k_\perp \to 0$. The terms in class $i)$ are not singular in the soft limit
and their contribution
in the collinear limit can thus be taken into account by supplementing the 
results
of this section with the ${\cal O}(\as)$-collinear factorization discussed in
Sect.~\ref{notations}.
Analogously, the terms in class $ii)$ are not 
singular in the collinear limit and their contribution in the soft limit can be 
taken into account by supplementing the results
of this section with the soft-gluon factorization
formula at ${\cal O}(\as)$ presented in Sect.~\ref{secsoftlo}.

This comment can be summarized in a formal manner by writing the square
of the matrix element as 
\beq
| {\cal M}_{g,a_1,a_2,\dots,a_n}(q,p_1,p_2,\dots,p_n) |^2 = 
\frac{1}{s_{12} s_{1q} s_{2q}} \;F(q,p_1,p_2,\dots,p_n) \;\;.
\eeq
The first factor on the right-hand side contains the correct scaling
behaviour in the soft and collinear regions. Thus, the soft--collinear
limit is defined by the soft $(q \to 0)$  and collinear $(k_\perp \to 0)$
approximations of the function $F(q,p_1,p_2,\dots,p_n)$ at
fixed ratio $q/k_\perp^2$.

To compute the soft--collinear limit we perform first soft approximations and
then collinear approximations.

The singular behaviour of the matrix element in the soft limit (and at fixed
$q/k_\perp^2$) is given by a factorization formula analogous to 
Eq.~(\ref{eikfac}), namely
\beq
\label{coleikfac}
\la c; \mu |\,{\cal M}_{g,a_1,a_2,\dots,a_n}(q,p_1,p_2,\dots,p_n) \rangle 
\simeq g_S \mu^\ep
J_{(12)}^{c;\mu}(q)
\; |\,{\cal M}_{a_1,a_2,\dots,a_n}(p_1,p_2,\dots,p_n)\rangle \;,
\eeq
but now the soft current $J_{(12)}^{c;\mu}(q)$ is no longer equal to the 
eikonal
current in Eq.~({\ref{eikcur}). In fact, since $p_1$ and $p_2$ can become
collinear, the internal partonic line with momentum $p_1+p_2$ in
${\cal M}_{a_1,a_2,\dots,a_n}(p_1,p_2,\dots,p_n)$ is close to the mass shell
$( \,(p_1+p_2)^2 = s_{12} \to 0)$. Near the mass shell, soft-gluon radiation
from this internal line leads to soft singularities and it cannot be neglected.

\begin{figure}[htb]
\begin{center}
\begin{tabular}{c}
\epsfxsize=10truecm
\epsffile{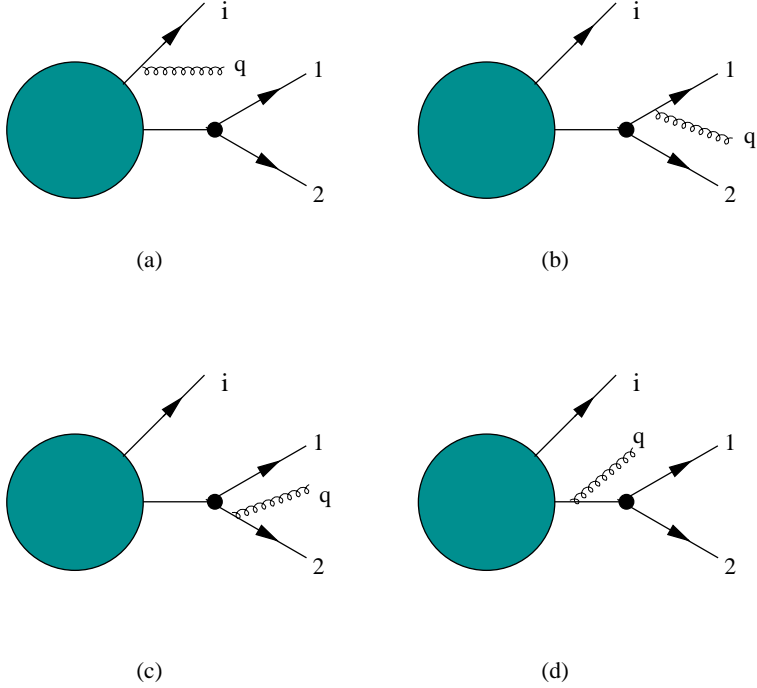}\\
\end{tabular}
\end{center}
\caption{\label{figsc}{\em Soft-gluon insertion diagrams for the soft--collinear
limit.
}}
\end{figure}

The explicit expression of the gluon current $J_{(12)}^{c;\mu}(q)$
can be derived by working in a physical gauge and
using the soft-gluon insertion rules described in
Sect.~\ref{secsoftlo} (Fig.~\ref{figsc}). We find
\beeq
\label{eikcolcur}
{\bom J}_{(12)}^\mu(q) &=& \sum_{i=3}^{n} {\bom T}_i\,\f{p_i^\mu}{p_i\cdot q} 
+ \frac{(p_1+p_2)^2}{(p_1+p_2+q)^2} 
\left[ {\bom T}_1\,\f{p_1^\mu}{p_1 \cdot q} 
+{\bom T}_2\,\f{p_2^\mu}{p_2 \cdot q} \right] \nonumber \\
&+& \left( {\bom T}_1 + {\bom T}_2 \right) \f{2(p_1^\mu+p_2^\mu)}{(p_1+p_2+q)^2} 
\;.
\eeeq
We discuss the three contributions on the right-hand side in turn.
The first contribution comes from the usual
eikonal insertions on the external parton lines $i=3,\dots,n$ (the diagrams
$(\rm a)$ in Fig.~\ref{figsc}).

The second contribution comes from the soft-gluon
emission from the external partons $p_1$ and $p_2$ (diagrams
$(\rm b)$ and $(\rm c)$ in Fig.~\ref{figsc}). The factor in the square bracket 
is the usual 
contribution from the eikonal vertices and propagators of the
lines $p_1+q$ and $p_2+q$. The factor in front of the square bracket
has the following origin. In Eq.~(\ref{coleikfac}) we have already factorized 
the tree amplitude ${\cal M}_{a_1,a_2,\dots,a_n}(p_1,p_2,\dots,p_n)$, which
contains the propagator factor $1/(p_1+p_2)^2$. In diagrams $(\rm b)$ and 
$(\rm c)$ of 
Fig.~\ref{figsc}, this propagator is instead absent, and it is replaced by the 
propagator $1/(p_1+p_2+q)^2$ of the internal line with momentum $p_1+p_2+q$.
Thus, the rescaling propagator factor $(p_1+p_2)^2/(p_1+p_2+q)^2$ has
to be applied to the contribution to the current.

The third contribution is the eikonal factor due to the soft emission
from the nearly on-shell internal line $p_1+p_2+q$ (diagram
$(\rm d)$ in Fig.~\ref{figsc}).

Note that we have neglected diagrams in which $p_1$ and $p_2$ are not produced
by a single line with momentum $p_1+p_2$.
These diagrams are not collinearly singular (see the discussion in 
Sect.~\ref{power}) in the physical gauge we are working on.

Note also that, as the eikonal current in Eq.~(\ref{eikcons}), the
soft current in Eq.~(\ref{eikcolcur}) satisfies the property
$q_\mu  {\bom J}_{(12)}^\mu(q) = \sum_{i=1}^{n} {\bom T}_i$. The ensuing
current conservation, which follows from Eq.~(\ref{cocon}), guarantees
the gauge invariance of the factorization formula (\ref{coleikfac}). 

Using Eq.~(\ref{coleikfac}) we could now perform the collinear limit
of the tree-level matrix element 
${\cal M}_{a_1,a_2,\dots,a_n}(p_1,p_2,\dots,p_n)$
on the right-hand side. However, since we are eventually interested
in the soft--collinear limit of the square of the matrix element
${\cal M}_{g,a_1,\dots,a_n}(q,p_1,\dots,p_n)$,
this procedure is not convenient for two reasons. First, we have to introduce 
collinear splitting functions for the various colour subamplitudes that
contribute to the colour vector
$|\,{\cal M}_{a_1,a_2,\dots,a_n}(p_1,p_2,\dots,p_n)\rangle$. These 
splitting functions differ from the Altarelli--Parisi splitting functions
of Sect.~\ref{notations}
(roughly speaking, the former are the square root
of the latter) and, although they are well known [\ref{mangano}], 
we shall show that they are not really necessary for the final result.
Secondly, the colour-charge transformation produced by the soft current
in Eq.~(\ref{coleikfac}) implies a non-trivial relation between the 
colour-subamplitude decomposition of the matrix element on the left-hand side
and the corresponding decomposition for the matrix element on the right-hand 
side. This non-trivial relation complicates the colour structure and
leads to mixed soft--collinear splitting functions [\ref{glover}], whose 
introduction can instead be avoided or, at least, simplified.

In other words, if we square the right-hand side of Eq.~(\ref{coleikfac}),
the soft current ${\bom J}_{(12)}$ produces non-trivial colour correlations
of the type ${\bom T}_1 \cdot {\bom T}_i$ or ${\bom T}_2 \cdot {\bom T}_i$
(with $i=3,\dots,n$) between 
${\cal M}(p_1,p_2,\dots,p_n)$ and ${\cal M}^\dagger(p_1,p_2,\dots,p_n)$.
Thus, we cannot perform the collinear limit $k_\perp \to 0$ by simply using the
known ${\cal O}(\as)$ results of Sect.~\ref{notations}
for the colour-summed
squared amplitude $|{\cal M}(p_1,p_2,\dots,p_n)|^2$.

The whole procedure can be simplified by exploiting the QCD {\em coherence}
properties of soft-gluon emission. We rewrite Eq.~(\ref{eikcolcur})
by splitting the soft current in two terms as follows:
\beq
\label{eikcoh}
{\bom J}_{(12)}^\mu(q) = 
\sum_{i=3}^{n} {\bom T}_i\,\f{p_i^\mu}{p_i\cdot q}
+ \left( {\bom T}_1 + {\bom T}_2 \right) \f{p_1^\mu+p_2^\mu}{(p_1+p_2) \cdot q}
+ \delta {\bom J}_{(12)}^\mu(q) \;,
\eeq
where
\beq
\label{delj}
\delta {\bom J}_{(12)}^\mu(q) = 
 \frac{(p_1+p_2)^2}{(p_1+p_2+q)^2} 
\left[ {\bom T}_1\,\f{p_1^\mu}{p_1 \cdot q} 
+{\bom T}_2\,\f{p_2^\mu}{p_2 \cdot q}  
- \left( {\bom T}_1 + {\bom T}_2 \right) \f{p_1^\mu+p_2^\mu}{(p_1+p_2) \cdot q}
\right] \;.
\eeq
The two terms, $\delta {\bom J}_{(12)}$ and the other contribution 
on the right-hand side of Eq.~(\ref{eikcoh}), are separately conserved
and, thus, the decomposition in Eq.~(\ref{eikcoh}) does not spoil the gauge
invariance.

Then we note that each of the two terms in Eq.~(\ref{eikcoh}) has the correct
scaling behaviour of ${\cal O}(1/q)$ when $q \to 0$. Their collinear behaviour
is nonetheless quite different. Performing the limit $k_\perp \to 0$ at fixed
$k_\perp^2/q$ in Eq.~(\ref{delj}), the propagator factor 
$(p_1+p_2)^2/(p_1+p_2+q)^2$ is of ${\cal O}(1)$ but the term in the
square bracket is of ${\cal O}(k_\perp/q)$. Thus, the contribution of
$\delta {\bom J}_{(12)}$ to the soft current ${\bom J}_{(12)}$ is suppressed
by a relative factor of ${\cal O}(k_\perp)$ in the collinear region, and
it can be neglected in the soft--collinear limit.

We conclude that in the factorization formula (\ref{coleikfac}) we can 
consistently use the following approximation for the soft current
in Eq.~(\ref{eikcolcur}):
\beq
\label{eikcohfin}
{\bom J}_{(12)}^\mu(q) \simeq \sum_{i=3}^{n} {\bom T}_i\,\f{p_i^\mu}{p_i\cdot q}
+ {\bom T}_{(12)} \f{p_1^\mu+p_2^\mu}{(p_1+p_2) \cdot q} \;\;,
\eeq 
where ${\bom T}_{(12)} = {\bom T}_1 + {\bom T}_2$. The subdominant effect of
$\delta {\bom J}_{(12)}$ is due to the cancellation between the different 
contributions
in the square bracket on the right-hand side of Eq.~(\ref{delj}). The 
cancellation is a typical consequence of colour coherence. When the parton
momenta $p_1$ and $p_2$ become collinear, they radiate soft gluon in 
a coherent way, i.e. as a single parton with momentum $p_1+p_2$ 
and colour charge ${\bom T}_{(12)} = {\bom T}_1 + {\bom T}_2$ (see the last
term in Eq.~(\ref{eikcohfin})).

The expression in Eq.~(\ref{eikcohfin}) is certainly simpler than that in 
Eq.~(\ref{eikcolcur}). More importantly, it depends on the colour charge
${\bom T}_{(12)}$  rather than separately on the colour charges
${\bom T}_1$ and ${\bom T}_2$. This implies that, when we square the amplitude 
in Eq.~(\ref{coleikfac}), the partons $p_1$ and $p_2$ are no longer 
colour-correlated, and the collinear limit $k_\perp \to 0$ can be performed by
using the collinear factorization formula (\ref{cfac}). We obtain the final
soft--collinear factorization formula:
\beeq
\label{scfac}
&&\!\!\!\!\!\!\!\!\!\! \!\!\!\!\!\!\!\!
| \cm_{g,a_1,a_2,\dots,a_n}(q,p_1,p_2,\dots,p_n) |^2 \simeq
- \frac{2}{s_{12}} \; (4 \pi \mu^{2\ep} \as)^2 \nn \\
&& \;\;\; \;\;\;\cdot \;
 \langle \,{\cal M}_{a,\dots,a_n}(p,\dots,p_n) \,|
\, {\hat {\bom P}}_{a_1 a_2} \, 
\left[{\bom J}_{(12) \mu}^{\dagger}(q) \,{\bom J}_{(12)}^{\mu}(q) \right]
| \, {\cal M}_{a,\dots,a_n}(p,\dots,p_n) \,\rangle 
\;, 
\eeeq
where the matrix elements on the right-hand side are obtained by removing
the soft gluon $q$ and by replacing the partons $a_1$ and $a_2$ by the single
parton $a$ that leads to the collinear splitting process $a \to  a_1 + a_2$.
Since these matrix elements are vectors in the colour+helicity space,
both spin and colour correlations are present in Eq.~(\ref{scfac}). 

The spin correlations are exactly the same as in Eq.~(\ref{cfac}).
The spin indices $s, s'$ of the parent parton $a$ are correlated by the
Altarelli--Parisi splitting functions
${\hat {\bom P}}_{a_1 a_2} \equiv {\hat P}_{a_1 a_2}^{ss'}(z,k_\perp;\ep)$ 
in Eqs.~(\ref{hpqqep})--(\ref{hpggep}).

The colour correlations affect all the partons and are
analogous to those in Eq.~(\ref{ccfact}). They
are produced by the square of the soft current:
\newpage
\beeq
\label{j12s}
{\bom J}_{(12) \mu}^{\dagger}(q) \,{\bom J}_{(12)}^{\mu}(q) &=&
\sum_{i,j=3}^n \;{\bom T}_i \cdot {\bom T}_j 
\;\frac{p_i \cdot p_j}{(p_i \cdot q) (p_j \cdot q)}
+ 2 \sum_{i=3}^n \; {\bom T}_i \cdot {\bom T}_{(12)}
\;\frac{p_i \cdot (p_1+p_2)}{(p_i \cdot q) (p_1+p_2) \cdot q} \nn \\
&+& {\bom T}_{(12)}^2 \;\frac{(p_1+p_2)^2}{((p_1+p_2)\cdot q)^2} \\
\label{j12sapp}
&\simeq& \sum_{i,j=3}^n \;{\bom T}_i \cdot {\bom T}_j \;{\cal S}_{ij}(q)
+ 2 \sum_{i=3}^n \; {\bom T}_i \cdot {\bom T}_{(12)}
\;{\cal S}_{i \,(12)}(q) \;\;.
\eeeq
In Eq.~(\ref{j12sapp}) we have neglected the last term on the right-hand side
of Eq.~(\ref{j12s}), because it is not collinearly singular.
We have also introduced the eikonal functions ${\cal S}_{ij}(q)$
of Eq.~(\ref{eikfun}) and the analogous eikonal function 
${\cal S}_{i \,(12)}(q)$,
\beq
\label{eikfunc}
{\cal S}_{i \,(12)}(q) = \frac{2 (s_{i1}+ s_{i2})}{s_{iq} (s_{1q}+ s_{2q})}
\;\;.
\eeq
From Eqs.~(\ref{scfac}) and (\ref{j12sapp}) we can see that the soft--collinear
limit at ${\cal O}(\as^2)$ is simply and fully described in terms of the same
factors, namely, soft eikonal functions and Altarelli--Parisi splitting 
functions, which control the soft and collinear limits at ${\cal O}(\as)$,
respectively.

The soft--collinear limit at ${\cal O}(\as^2)$ was first studied by Campbell
and Glover [\ref{glover}]. They neglected spin correlations and considered
the singular behaviour of the colour subamplitudes. This behaviour, which was 
extracted by directly performing the singular limit of known squared matrix
elements, was given in terms of two different factors. The first factor
(see Sect.~4.4 in Ref.~[\ref{glover}]) refers to subamplitudes in which the
collinear partons are not colour-connected and it corresponds exactly  to the
${\cal S}_{ij}(q)$-term in Eq.~(\ref{j12sapp}). The second factor regards
the subamplitudes in which the collinear partons $p_1$ and $p_2$ are
colour-connected. This factor is given in Sect.~5.2 of Ref.~[\ref{glover}]
and it can be written as 
\beeq
S_{i; \,q12} &=& \f{2(s_{i1}+s_{i2})}{s_{iq} \;s_{1q}}
\left[ z + \f{s_{1q} + zs_{12}}{s_{12q}} \right] \; \\
\label{sscid}
&=&\f{2(s_{i1}+s_{i2})}{s_{iq}}
\left\{ \f{2}{s_{1q}+s_{2q}}+\left(1+\f{s_{12}}{s_{12q}}\right)
\left[\f{z}{s_{1q}}-\f{1}{s_{1q}+s_{2q}}\right]\right\} \\
\label{sscapp}
&\simeq& \f{4(s_{i1}+s_{i2})}{s_{iq} (s_{1q}+s_{2q})} \;\;.
\eeeq
Since $z$ is the longitudinal momentum fraction carried by $p_1$ in the
collinear region, the term in the square bracket of
Eq.~(\ref{sscid}) vanishes in the collinear limit and Eq.~(\ref{sscapp}) 
follows. This simplification, which is due to colour coherence, was not
performed in Ref.~[\ref{glover}]. Taking it into account, we have
$S_{i;\,q12} \simeq 2 {\cal S}_{i \,(12)}(q)$, which, when inserted in 
Eq.~(\ref{j12sapp}), shows the equivalence of our results with those of
Ref.~[\ref{glover}].

Our derivation of the soft--collinear factorization formula (\ref{scfac})
can straightforwardly be extended to higher orders. We can consider  
the limit where a single gluon with momentum $q$ becomes soft and,
at the same time, $m$ partons, say $p_1, \dots , p_m$,
become simultaneously collinear (see Sect.~\ref{power}). In this limit the
factorization formula is
\beeq
\label{smcfac}
&&\!\!\!\!\!\!\!\!\!\! \!\!\!\!\!\!\!\!
| \cm_{g,a_1,\dots,a_m,\dots,a_n}(q,p_1,\dots,p_m,\dots,p_n) |^2 \simeq
- \;4 \pi \mu^{2\ep} \as 
\left( \frac{8 \pi \mu^{2\ep} \as}{s_{1 \dots m}}\right)^{m-1}
\nn \\
&& \!\!\! \!\!\!\cdot \;
 \langle \,{\cal M}_{a,\dots,a_n}(xp,\dots,p_n) \,|
\, {\hat {\bom P}}_{a_1 \dots a_m} \, 
\left[{\bom J}_{(1 \dots m) \mu}^{\dagger}(q) 
\,{\bom J}_{(1 \dots m)}^{\mu}(q) \right]
| \, {\cal M}_{a,\dots,a_n}(xp,\dots,p_n) \,\rangle 
\;, 
\eeeq
where ${\hat {\bom P}}_{a_1 \dots a_m} \equiv 
{\hat P}_{a_1 \dots a_m}^{s s'}$ is the spin-dependent splitting 
function in Eq.~(\ref{ccfacm}), and the soft current 
${\bom J}_{(1 \dots m)}(q)$ is: 
\beq
\label{eikmcohfin}
{\bom J}_{(1 \dots m)}^\mu(q) \simeq 
\sum_{i=m+1}^{n} {\bom T}_i\,\f{p_i^\mu}{p_i\cdot q}
+ {\bom T}_{(1 \dots m)} \f{p_1^\mu+\dots p_m^\mu}{(p_1+\dots +p_m) \cdot q} 
\;\;,
\eeq 
with ${\bom T}_{(1 \dots m)} = {\bom T}_1 + \dots + {\bom T}_m$.
Squaring the soft current as in Eqs.~(\ref{j12s}) and (\ref{j12sapp}), we obtain
\beq
\label{j1ms}
{\bom J}_{(1 \dots m) \mu}^{\dagger}(q) \,{\bom J}_{(1 \dots m)}^{\mu}(q) 
\label{j12smapp}
\simeq \sum_{i,j=m+1}^n \;{\bom T}_i \cdot {\bom T}_j \;{\cal
  S}_{ij}(q)
+ 2 \sum_{i=m+1}^n \; {\bom T}_i \cdot {\bom T}_{(1 \dots m)}
\;{\cal S}_{i \,(1 \dots m)}(q) \;\;,
\eeq
where ${\cal S}_{i \,(1 \dots m)}(q) =
2 (s_{i1}+ \dots + s_{im})/[ s_{iq} (s_{1q}+ \dots + s_{mq})]$.

The proof of these results is very simple.
The soft current ${\bom J}_{(1 \dots m)}(q)$ is derived by using the soft-gluon
insertion rules as in Eq.~(\ref{eikcolcur}). Then, the coherence argument
used in Eqs.~(\ref{eikcoh}) and (\ref{delj}) can iteratively be applied to 
any vertex in the $m$-parton dispersive amplitude
${\cal V}_{a_1 \dots a_m}$ of Eq.~(\ref{fcollgen}). This leads to the 
expression in Eq.~(\ref{eikmcohfin}).

\subsection{Multiple soft and soft--collinear limits}
\label{multilim}

In Sects.~(\ref{secsoftlo}) and (\ref{secsoftgg}) we have discussed in 
detail single and double soft-gluon emission. The generalization to multiple
soft-gluon radiation is straightforward. If we consider the matrix
element $\cm_{g,\dots,g,a_1,\dots,a_n}(q_1,\dots,q_k,p_1,\dots,p_n)$
when the $k$ gluons with momenta $q_1,\dots,q_k$ become soft
simultaneously, we can still write a factorization formula similar to 
Eq.~(\ref{softff2}) by performing the simple replacement
\beq
\label{softgen}
g_S^2 \mu^{2\ep} \, J^{a_1a_2}_{\mu_1\mu_2}(q_1,q_2)
\to
\left( g_S \mu^{\ep} \right)^k \, 
J^{a_1\dots a_k}_{\mu_1\dots \mu_k}(q_1,\dots,q_k) \;\;,
\eeq
where $a_1, \dots, a_k$ and $\mu_1, \dots, \mu_k$ denote the colour and 
Lorentz indices of the soft gluons.
As the two-gluon current $J^{a_1a_2}_{\mu_1\mu_2}(q_1,q_2)$
in Eq.~(\ref{dsoftcur}), the multigluon current on the right-hand side of
Eq.~(\ref{softgen}) is obtained by working in a physical gauge and using
the soft-gluon insertion rules described in Sect.~\ref{secsoftlo}. Of course,
the explicit expression of ${\bom J}(q_1,\dots,q_k)$ turns out to be quite 
involved, because all the possible interactions between the soft gluons have to 
be included without using any soft approximation.

It is also clear that the soft-gluon insertion rules can be used to
derive a factorization formula analogous to Eq.~(\ref{coleikfac}) for the 
multiple soft--collinear limit in which $k$ gluons are soft and 
$m$ partons become collinear simultaneously. More importantly, it is worth while
pointing out that the {\em coherence} argument leading to Eqs.~(\ref{scfac})
and (\ref{smcfac}) still applies.
Thus, the factorization formula can be written as
\beeq
\label{scfacgen}
&&\!\!\!\!\!\!\!\!\!\! 
| \cm_{g,\dots,g,a_1,\dots,a_m,\dots,a_n}(q_1,\dots,q_k,p_1,\dots,p_m,
\dots,p_n) |^2 \simeq
\left( - 4 \pi \mu^{2\ep} \as\right)^k 
\left( \frac{8 \pi \mu^{2\ep} \as}{s_{1 \dots m}}\right)^{m-1}
\\
&& \!\!\! \!\!\!\!\!\!\!\cdot \;
 \langle \,{\cal M}_{a,\dots,a_n}(xp,..,p_n) \,|
\, {\hat {\bom P}}_{a_1 \dots a_m} \, 
\left[{\bom J}_{(1 \dots m)}^{\dagger}(q_1,\dots,q_k) 
\,{\bom J}_{(1 \dots m)}(q_1,\dots,q_k) \right]
| \, {\cal M}_{a,\dots,a_n}(xp,..,p_n) \,\rangle \nn
\;. 
\eeeq
Equation (\ref{scfacgen}) does not involve any additional factor with respect
to those that are necessary to deal with the multiple collinear and multiple
soft limits separately. The spin-dependent splitting function
${\hat {\bom P}}_{a_1 \dots a_m} \equiv {\hat P}_{a_1 \dots a_m}^{s s'}$
is exactly the same as that in Eq.~(\ref{ccfacm}). The current
${\bom J}_{(1 \dots m)}(q_1,\dots,q_k)$ is completely analogous to the soft
current ${\bom J}(q_1,\dots,q_k)$ in Eq.~(\ref{softgen}). As the latter,
the former is constructed by inserting the soft gluons only on the
$(1+n-m)$ {\em external} parton lines with momenta
$p_1+\dots+p_m, p_{m+1}, \dots, p_n$, and each insertion on the collinear 
parton $p_1+\dots+p_m$ is taken into account by the simple eikonal factor 
$({\bom T}_1 + \dots + {\bom T}_m) (p_1+\dots+p_m)^\mu/(p_1+\dots+p_m)\cdot q$,
despite the fact that $(p_1+\dots+p_m)^2 \neq 0$ (see e.g. 
Eqs.~(\ref{eikmcohfin})).

In the most general case, the infrared singularities of the tree-level 
amplitudes
are produced by the multiple collinear decay of hard partons and by the 
associated radiation of soft gluons and $q{\bar q}$-pairs. The corresponding 
factorization formula can
be constructed in a straightforward manner by using the rules derived and 
illustrated throughout the paper. Using a shorthand symbolic notation, we 
have~(Fig.~\ref{general})
\beq
|\cm|^2\simeq \langle \cm_{hard}|
\, \left(\prod_i {\hat {\bom P}}_i\right) \,{\bom S}\, |\cm_{hard}\rangle \;\;.
\eeq
Here $\cm_{hard}$ denotes the factorized amplitude that depends only on the
momenta of the hard partons. The factor $\prod_i {\hat {\bom P}}_i=\prod_i 
{\hat P}_i^{s_is_i'}$ is the product of the spin-dependent splitting functions 
for the collinear decay of $i=1,\dots,l$ hard partons. The factor ${\bom S}$ is 
a colour matrix that takes into account the radiation of soft partons. It has 
to be computed exactly at the tree level, but its external {\em gluon} lines 
are coupled to the hard partons by using the eikonal approximation as in the 
calculation of the current 
${\bom J}_{(1 \dots m)}(q_1,\dots,q_k)$ in Eq.~(\ref{scfacgen}). Note that spin
and colour correlations are factorized independently. This decoupling follows 
from colour coherence.

\begin{figure}[htb]
\begin{center}
\begin{tabular}{c}
\epsfxsize=10truecm
\epsffile{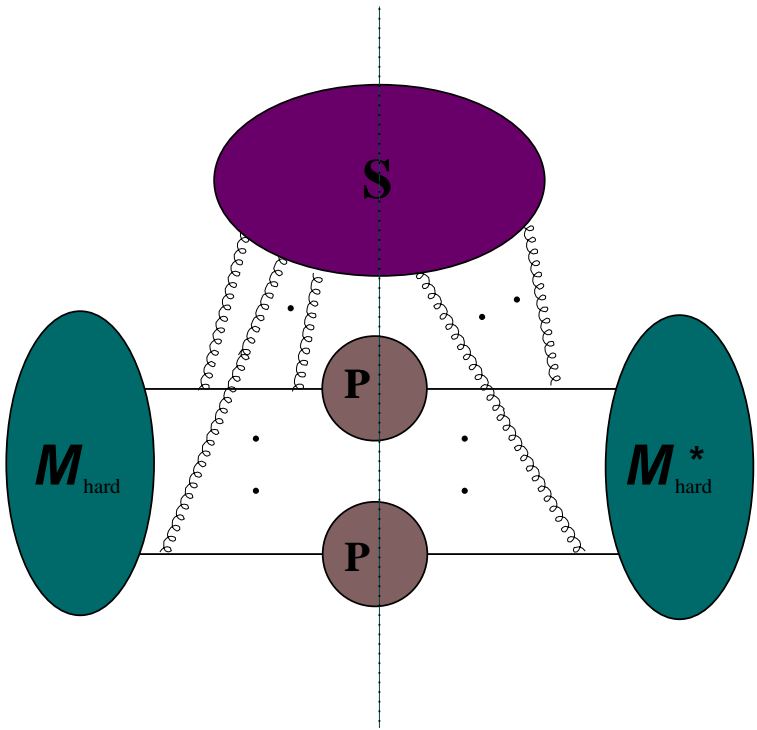}\\
\end{tabular}
\end{center}
\caption{\label{general}{\em General structure of infrared factorization
at any perturbative order.
}}
\end{figure}

\section{Summary}
\label{summa}

In this paper we have studied the infrared structure of tree-level QCD
amplitudes in all the possible soft and collinear limits.

We have first considered the collinear behaviour.
We have shown that, in the limit in which $m$ partons become parallel
simultaneously, the singularities are given by
the universal factorization formula (\ref{ccfacm}) and are controlled by
process-independent splitting functions that generalize the customary
Altarelli--Parisi splitting functions. These splitting functions fully 
take into account the azimuthal correlations produced in the collinear 
decay. We have presented a recipe to compute the splitting functions at 
any perturbative order and we have performed their explicit calculation 
at ${\cal O}(\as^2)$.

Then we have studied the soft behaviour and shown how to construct
soft factorization formulae at any order in $\as$. We have considered 
the limit in which a $q{\bar q}$ pair becomes soft and we have computed 
the corresponding singularity at ${\cal O}(\as^2)$ in terms of a simple 
universal insertion factor. We have then recalled the known results about
the limit in which two gluons become soft. This limit is controlled by an 
${\cal O}(\as^2)$ soft current that is tensor in colour space and
generalizes the eikonal current at ${\cal O}(\as)$. We have obtained
a compact expression for the square of the two-gluon current that, in
particular, shows the absence of colour correlations in the case of
four- and five-parton amplitudes.

Finally, we have studied the mixed soft--collinear limit and pointed out
that its description does not require the introduction of new infrared
factors. Exploiting the coherence property of soft gluon radiation, 
we have been able to show that 
the singularities are given by a factorization formula written only in terms 
of the soft currents and of the splitting functions that control the soft
and collinear limits, respectively.

These results are one of the necessary ingredients to extend QCD predictions 
at higher perturbative orders. In particular, our calculation of the
${\cal O}(\as^2)$ singular factors is relevant to setting up general methods to
compute QCD jet cross sections at NNLO.

\noindent {\bf Acknowledgements}. \\
\noindent 
We would like to thank Daniel de Florian and Zoltan Kunszt for discussions.
One of us (M.G.) would like to thank the 
Fondazione `Angelo della Riccia' and the INFN for financial support at 
earlier stages of this work.


\renewcommand{\theequation}{A.\arabic{equation}}
\setcounter{equation}{0}

\section*{Appendix: Soft limits of four- and five-parton amplitudes}
\label{appendixa}

In general,
soft factorization formulae involve colour correlations. 
As shown in Eqs.~(\ref{qqsoftfac}), (\ref{dsoftm2}) and (\ref{j12sapp}), at 
${\cal O}(\as^2)$ the correlations are completely given in terms of
products of colour-charge factors ${\bom T}_i\cdot{\bom T}_j$.

In this appendix we collect the factorization formulae for the ${\cal O}(\as^2)$
soft (and soft--collinear) limits of the square of the
matrix elements with four and five partons plus an arbitrary number of 
colourless particles. In these particular cases, using 
colour conservation, it is possible (see e.g. the Appendix~A in 
Ref.~[\ref{CSdipole}]\, ) to express the products 
${\bom T}_i\cdot{\bom T}_j$ in terms of the Casimir invariants $C_i$
($C_i=C_F$ if $i=q,{\bar q}$ and $C_i=C_A$ if $i=g$)
of the hard partons. Thus the colour algebra completely
factorizes and colour correlations cancel.

Note that two of the hard partons in the four- and five-parton amplitudes
necessarily form a particle--antiparticle pair $a, {\bar a}$. This further
simplifies the combinations of Casimir invariants that appear in the
factorization formulae.

\noindent {\bf Emission of a soft $q{\bar q}$-pair}

We consider the four-parton amplitude 
${\cal M}_{q,{\bar q},a,{\bar a}}(q_1,q_2,p_1,p_2)$ in the limit in 
which 
\newline
$q_1,q_2\to~0$. From Eq.~(\ref{qqsoftfac}), we get
\beq
\label{qq4}
\!\!\!\!\!\! |{\cal M}_{q,{\bar q},a,{\bar a}}\,(q_1,q_2,p_1,p_2)|^2 \simeq
(4 \pi\mu^{2\ep} \as)^2\, T_R\; C_a
\; \Bigg( \;{\cal I}_{11} +
{\cal I}_{22} -2\,{\cal I}_{12} \,\Bigg) \;
|{\cal M}_{a,{\bar a}}\,(p_1,p_2)|^2 \;.
\eeq
In the case of five partons we get
\beeq
\label{qq5}
|{\cal M}_{q,{\bar q},a,{\bar a},a_3}\,(q_1,q_2,p_1,p_2,p_3)|^2
&\simeq& (4 \pi\mu^{2\ep} \as)^2\,T_R\;
\Bigg[C_a \left( \,{\cal I}_{11}+{\cal I}_{22} - 2 {\cal I}_{12} \right)  \\
&+& C_{a_3} \left( \,{\cal I}_{33} + {\cal I}_{12} - {\cal I}_{13} 
- {\cal I}_{23}\right) \Bigg] 
\;|{\cal M}_{a,{\bar a},a_3}\,(p_1,p_2,p_3)|^2 \,. \nn
\eeeq
The soft function ${\cal I}_{ij}={\cal I}_{ij}(q_1,q_2)$ is given in 
Eq.~(\ref{Iij1}).

\noindent {\bf Emission of two soft gluons}

We consider the amplitude ${\cal M}_{g,g,a,{\bar a}}\,(q_1,q_2,p_1,p_2)$ in
the limit in which the two gluons become soft. Using Eq.~(\ref{dsoftm2}),
we get
\beeq
\label{gg4}
|{\cal M}_{g,g,a,{\bar a}}\,(q_1,q_2,p_1,p_2)|^2 \!&\simeq&\!\! 
(4 \pi \mu^{2\ep} \as)^2\, C_a\; \Bigg[4\,C_a\,{\cal S}_{12}(q_1)
\,{\cal S}_{12}(q_2) + C_A\,\Big( 2\,{\cal S}_{12} - {\cal S}_{11}
-{\cal S}_{22} \Big)\Bigg]\nn\\
&\cdot& |{\cal M}_{a,{\bar a}}\,(p_1,p_2)|^2 \,.
\eeeq
In the case of five partons we get
\beeq
\label{gg5}
&&\!\!\!\!\! \!\!\!\!\!\!\!\!\!\! \!\!\!\!\!\!\!\!\!\!\!\!\!\!\!\!\!\!\!\!
\!\!\!\!\!
|{\cal M}_{g,g,a,{\bar a},a_3}\,(q_1,q_2,p_1,p_2,p_3)|^2 
\simeq (4 \pi \mu^{2\ep} \as)^2 \;
|{\cal M}_{a,{\bar a},a_3}\,(p_1,p_2,p_3)|^2 \nn\\
\;\;\;\;\;\;&\cdot&\!\!\!\!\! \Bigg\{ \;\Bigg[ 
\left( 2 C_a - C_{a_3} \right) {\cal S}_{12}(q_1)
+ C_{a_3} \left( {\cal S}_{13}(q_1) + {\cal S}_{23}(q_1) 
\right) \Bigg] \nn\\
\;\;\;\;\;\;\;\;\; \;\;&\cdot& 
\Bigg[ \left( 2 C_a - C_{a_3} \right) {\cal S}_{12}(q_2)
+ C_{a_3} \left( {\cal S}_{13}(q_2) + {\cal S}_{23}(q_2)
\right) \Bigg] \\
\;\;\;\;\;\;&+&\!\!\!C_A\,\Bigg[\,C_a \left( 2 {\cal S}_{12} - {\cal S}_{11}
-{\cal S}_{22} \right) + C_{a_3} \left( {\cal S}_{13}+{\cal S}_{23}
- {\cal S}_{33} - {\cal S}_{12}
\right) \Bigg] \Bigg\}\,. \nn
\eeeq
The soft functions ${\cal S}_{ij}(q)$ and ${\cal S}_{ij}={\cal S}_{ij}(q_1,q_2)$
are given in Eqs.~(\ref{eikfun}) and (\ref{dsoftfun}), respectively.

\newpage

\noindent {\bf Soft--collinear limit}

We consider the amplitude ${\cal M}_{g,a_1,a_2,a_3}\,(q,p_1,p_2,p_3)$ in the
limit in which $q\to 0$ and $s_{12}\to 0$. Using Eq.~(\ref{scfac}), we get
\beq
\label{sc4}
|{\cal M}_{g,a_1,a_2,a_3}(q,p_1,p_2,p_3)|^2
\simeq \f{4}{s_{12}} (4 \pi \mu^{2\ep} \as)^2\,C_{a_3}\,{\cal S}_{3(12)}(q)
\;{\hat P}_{a_1a_2}^{ss'} \;{\cal T}_{aa_3}^{ss'}(xp,p_3) \;\;.
\eeq
In the case of five partons we get
\beeq
\label{sc5}
&&\!\!\!\!\!\!\!\!\!\!\!\!\!\!\!\!\!\!\! 
|{\cal M}_{g,a_1,a_2,a_3,a_4}(q_1,p_1,p_2,p_3,p_4)|^2
\simeq \f{2}{s_{12}} (4 \pi \mu^{2\ep} \as)^2
\;{\cal T}_{aa_3a_4}^{ss'}(xp,p_3,p_4) \;{\hat P}^{ss'}_{a_1a_2} \\
&&\!\!\!\!\!\!\!\!\!\cdot \Bigg[ (C_{a_3}+C_{a_4}-C_{a})\,{\cal S}_{34}(q)
+(C_{a}+C_{a_3}-C_{a_4})\,{\cal S}_{3(12)}(q)
+(C_{a}+C_{a_4}-C_{a_3})\,{\cal S}_{4(12)}(q)\Bigg] \;. \nn
\eeeq
Here $a$ denotes the parton that decays collinearly, $a \to a_1 a_2$,
${\cal T}_{a \dots}^{ss'}(xp,\dots)$ is the spin-polarization tensor in
Eq.~(\ref{melspindef}) and ${\hat P}^{ss'}_{a_1a_2}$ is the spin-dependent 
splitting function in Eq.~(\ref{cfac}).
The soft functions ${\cal S}_{ij}(q)$ and ${\cal S}_{i(12)}(q)$
are given in Eqs.~(\ref{eikfun}) and (\ref{eikfunc}), respectively.

\section*{References}

\def\ac#1#2#3{Acta Phys.\ Polon.\ #1 (19#3) #2}
\def\ap#1#2#3{Ann.\ Phys.\ (NY) #1 (19#3) #2}
\def\ar#1#2#3{Annu.\ Rev.\ Nucl.\ Part.\ Sci.\ #1 (19#3) #2}
\def\cpc#1#2#3{Computer Phys.\ Comm.\ #1 (19#3) #2}
\def\ib#1#2#3{ibid.\ #1 (19#3) #2}
\def\np#1#2#3{Nucl.\ Phys.\ B#1 (19#3) #2}
\def\pl#1#2#3{Phys.\ Lett.\ #1B (19#3) #2}
\def\pr#1#2#3{Phys.\ Rev.\ D #1 (19#3) #2}
\def\prep#1#2#3{Phys.\ Rep.\ #1 (19#3) #2}
\def\prl#1#2#3{Phys.\ Rev.\ Lett.\ #1 (19#3) #2}
\def\rmp#1#2#3{Rev.\ Mod.\ Phys.\ #1 (19#3) #2}
\def\sj#1#2#3{Sov.\ J.\ Nucl.\ Phys.\ #1 (19#3) #2}
\def\zp#1#2#3{Z.\ Phys.\ C#1 (19#3) #2}

\begin{enumerate}


\item \label{book}
R.K.\ Ellis, W.J.\ Stirling and B.R.\ Webber, {\it QCD and collider 
physics} (Cambridge University Press, Cambridge, 1996) and references therein.

\item \label{AP}
G.\ Altarelli and G.\ Parisi, \np{126}{298}{77}.

\item \label{BCM}
A.\ Bassetto, M.\ Ciafaloni and G.\ Marchesini, \prep{100}{201}{83};
Yu.L.~Dokshitser, V.A.\ Khoze, A.H.\ Mueller and S.I. Troian,
{\it Basics of Perturbative QCD} (Editions Fronti\`eres, Gif-sur-Yvette, 1991)
and references therein.

\item \label{mangano}
M.L.\ Mangano and S.J.\ Parke, \prep{200}{301}{91}
and references therein.

\item \label{CSdipole}
S.\ Catani and M.H.\ Seymour, \np{485}{291}{97}
(Erratum {\it ibid.} B510 (1998) 503).

\item \label{antenna} 
D.A.\ Kosower, \pr{57}{5410}{98}.

\item \label{GG}
W.T. Giele and E.W.N. Glover, \pr{46}{1980}{92}.

\item \label{KST}
Z.\ Kunszt, A.\ Signer and Z. Tr\'ocs\'anyi, \np{420}{550}{94}.

\item \label{CSdipolelet}
S.\ Catani and M.H.\ Seymour, \pl{378}{287}{96}.

\item \label{BDKrev}
Z.\ Bern, L.\ Dixon and D.A.\ Kosower, \ar{46}{109}{96} and references therein.

\item \label{GGK}
W.T. Giele, E.W.N. Glover and D.A. Kosower, \np{403}{633}{93};
S.\ Keller and E.\ Laenen, \pr{59}{114004}{99}.

\item \label{submeth}
Z.\ Kunszt and D.E.\ Soper, \pr{46}{192}{92};
S.\ Frixione, Z.\ Kunszt and A.\ Signer, \np{467}{399}{96};
Z. Nagy and Z. Tr\'ocs\'anyi, \np{486}{189}{97};
S.\ Frixione, \np{507}{295}{97}.

\item \label{gonsalves}
R.J.\ Gonsalves, \pr{28}{1542}{83};
G.\ Kramer and B.\ Lampe, \zp{34}{497}{87} (Erratum {\it ibid.} C42 (1989) 504);
T.\ Matsuura, S.C.\ van der Marck and W.L.\ van Neerven, \np{319}{570}{89}.

\item \label{smirnov}
V.A. Smirnov, hep-ph/9905323; 
V.A.\ Smirnov and O.L.\ Veretin, preprint DESY-99-100 (hep-ph/9907385).  

\item \label{bern}
See, for instance, Z.\ Bern, J.S.\ Rozowsky and B.\ Yan, \pl{401}{273}{97};
C.\ Anastasiou, E.W.N.\ Glover and C.\ Oleari, preprint DTP/99/80
(hep-ph/9907494); and references therein.

\item \label{sing2loop}
S.\ Catani, \pl{427}{161}{98}.

\item \label{1loopeps}
Z.\ Bern, V.\ Del Duca and C.R.\ Schmidt, \pl{445}{168}{98};
Z.\ Bern, V.\ Del Duca, W.B. Kilgore and C.R.\ Schmidt, 
preprint BNL-HET-99-6 (hep-ph/9903516).

\item  \label{1loopepskos}
D.A.\ Kosower, \np{552}{319}{99}; 
D.A.\ Kosower and P.\ Uwer, preprint SACLAY-SPHT-T99-032
(hep-ph/9903515).

\item \label{bgdsoft}
F.A.\ Berends and W.T.\ Giele, \np{313}{595}{89}.

\item \label{sdsoft}
S.\ Catani, in Proceedings of the Workshop on {\it New Techniques for
Calculating Higher Order QCD Corrections}, report ETH-TH/93-01, Zurich (1992). 

\item \label{glover}
J.M.\ Campbell and E.W.N. Glover, \np{527}{264}{98}. 

\item \label{lett}
S.\ Catani and M.\ Grazzini, \pl{446}{143}{99}.

\item \label{collpc}
D.\ Amati, R.\ Petronzio and G.\ Veneziano, \np{140}{54}{78}, 
\np{146}{29}{78}; R.K.\ Ellis, H.\ Georgi, M.\ Machacek, H.D.\ Politzer and
G.G.\ Ross, \pl{78}{281}{78}, \np{152}{285}{79}.

\item \label{jetcalc}
J.\ Kalinowski, K.\ Konishi and T.R.\ Taylor, \np{181}{221}{81};
J.\ Kalinowski, K.\ Konishi, P.N.\ Scharbach and T.R.\ Taylor, 
\np{181}{253}{81}; J.F.\ Gunion, J.\ Kalinowski and L.\ Szymanowski,
\pr{32}{2303}{85}.

\item \label{coher}
B.I.\ Ermolaev and V.S.\ Fadin, JETP Lett. 33 (1981) 269.

\item \label{KUV}
K.\ Konishi, A.\ Ukawa and G.\ Veneziano, \np{157}{45}{79}.

\item \label{softrev}
G.\ Sterman, in Proc. {\it 10th Topical Workshop on Proton-Antiproton
Collider Physics}, eds. R.\ Raja and J.\ Yoh (AIP Press, New York, 1996),
p.~608; 
S.\ Catani, 
in Proc. of the {\it 32nd Rencontres de Moriond: QCD and High-Energy
Hadronic Interactions}, ed. J. Tran Than Van (Editions Fronti\`eres, Paris,
1997), p.~331 and references therein.

\item \label{cdruv}
G.\ 't Hooft and M.\ Veltman, \np{44}{189}{72};
G.\ Bollini and J.J. Giambia\-gi, Nuovo Cimento 12B (1972) 20;
J.F.\ Ashmore, Nuovo Cimento Lett. 4 (1972) 289;
G.M. Cicuta and E.\ Montaldi, Nuovo Cimento Lett. 4 (1972) 329.

\item \label{cdrir}
R.\ Gastmans and R.\ Meuldermans, \np{63}{277}{73}.

\item \label{schemedep}
S.\ Catani, M.H.\ Seymour and Z.\ Tr\'ocs\'anyi, \pr{55}{6819}{97}.

\item \label{dimred}
W.\ Siegel, \pl{84}{193}{79}; Z.\ Bern and D.A.\ Kosower, \np{379}{451}{92}.

\end{enumerate}

\end{document}